\documentclass[aps,prresearch,twocolumn,superscriptaddress,nofootinbib,longbibliography]{revtex4-2}

\usepackage{graphicx}
\usepackage{amsmath,amssymb,bm}
\usepackage{xcolor}
\usepackage{pifont}
\usepackage{tikz}
\usetikzlibrary{calc}
\usepackage{pgfplots}
\usepgfplotslibrary{groupplots}
\pgfplotsset{compat=1.18}
\usepackage[colorlinks=true,linkcolor=blue,citecolor=blue,urlcolor=blue]{hyperref}

\definecolor{cBlue}{RGB}{31,119,180}
\definecolor{cOrange}{RGB}{230,120,20}
\definecolor{cRed}{RGB}{200,35,35}
\definecolor{cGreen}{RGB}{34,130,60}
\definecolor{cPurple}{RGB}{125,90,180}
\definecolor{cTeal}{RGB}{0,128,128}
\definecolor{cGray}{RGB}{120,120,120}

\pgfplotsset{
  rrfig/.style={
    width=0.99\columnwidth, height=0.60\columnwidth,
    tick label style={font=\footnotesize},
    label style={font=\footnotesize},
    legend style={font=\scriptsize, draw=none, fill=none},
    line width=0.9pt, tick align=outside, tick pos=left,
    axis line style={line width=0.5pt},
    every axis plot/.append style={line join=round},
  }
}

\newcommand{\dd}{\mathrm{d}}

\begin{document}

\title{Motional refocusing for trap-off Rydberg gates}

\author{Yoav Sagi}

\email[Electronic address: ]{yoavsagi@technion.ac.il}

\affiliation{Physics Department and Solid State Institute, Technion - Israel Institute of Technology, Haifa 32000, Israel}
\affiliation{Q-Factor, Ramat Gan, Israel}

\author{Ofer Firstenberg}
\affiliation{Department of Physics of Complex Systems, Weizmann Institute of Science, Rehovot 7610001, Israel}
\affiliation{Q-Factor, Ramat Gan, Israel}

\author{Nir Davidson}
\affiliation{Department of Physics of Complex Systems, Weizmann Institute of Science, Rehovot 7610001, Israel}
\affiliation{Q-Factor, Ramat Gan, Israel}

\author{Guy Raz}
\affiliation{Q-Factor, Ramat Gan, Israel}

\date{\today}

\begin{abstract}

Rydberg entangling gates in optical-tweezer arrays are commonly executed with the trapping light switched off, so every gate contains a release--and--recapture cycle that heats the atomic motion and can ultimately limit circuit depth. We develop a motional refocusing protocol that exactly removes this heating in the harmonic approximation using only programmable intensity switching of the trapping light. The protocol closes the release--and--recapture cycle for every matched harmonic mode, returning arbitrary motional populations and coherences exactly up to ordinary evolution under the static trap. We derive the recovery sequence in closed form for arbitrary catch depth and prove that, within the experimentally relevant regime, it is the unique globally time-optimal solution under bounded trap intensity. The harmonic theory is then extended in two directions. First, we construct exact common-intensity recovery sequences that simultaneously refocus several nondegenerate harmonic modes, including radial--axial and fully anisotropic three-dimensional traps. Second, we derive a composite sequence that suppresses the leading anharmonic correction of weakly anharmonic traps by canceling all first-order motional transitions induced by the quartic anharmonicity, changing the residual heating law from $U_0^{-2}$ to $U_0^{-4}$. Wave-packet simulations in realistic Gaussian tweezers validate the analytic theory and quantify the residual effects of anharmonicity, finite switching ramps, trap ellipticity, and control errors. Applied to representative cesium Rydberg gates, the protocol suppresses the dominant recapture heating to the anharmonic floor and prevents the associated motional Doppler contribution from increasing with circuit depth. The resulting framework provides a practical route toward heating-free trap-off neutral-atom gates using only trap-intensity modulation.
\end{abstract}

\maketitle

\section{Introduction}
\label{sec:intro}

Arrays of single neutral atoms held in optical tweezers have become a leading platform for quantum computation and simulation~\cite{Kaufman2021,Browaeys2020,Saffman2010,Saffman2016}, with the Rydberg blockade gate as the native entangling operation~\cite{Jaksch2000,Levine2019,Evered2023,Bluvstein2024,Madjarov2020,Cao2024}. During the gate the tweezers are usually switched off, because the Rydberg state is anti-trapped by the ground-state trapping light~\cite{deLeseleuc2018,deKeijzer2023,Zhang2011,Barredo2020,Wilson2022} and the differential light shift would otherwise dephase the gate. Two-qubit gates therefore contain a release-and-recapture cycle, repeated at every layer of a deep circuit.

This cycle carries an accumulated thermodynamic cost. During the dark time $T$ the wave packet expands and acquires a position--momentum correlation; when the trap is suddenly restored, the state is no longer stationary, and the atom is heated. For an atom prepared near the motional ground state~\cite{Kaufman2012,Thompson2013}, recapture heating is typically the dominant per-gate motional-energy increase, it compounds geometrically with circuit depth once the breathing excitation dephases between gates, and it feeds directly into the motional (Doppler) contribution to the gate error budget~\cite{deLeseleuc2018,Robicheaux2021,Pagano2022,Jandura2022}. Current mitigations, such as ramping the trap depth down before release and back up after~\cite{Cao2024,Madjarov2020} or keeping the traps on at reduced depth~\cite{Wilson2022,Barredo2020}, trade heating against gate dephasing and add adiabatic time scales long compared with the inverse of the trapping frequency, $\omega^{-1}$. What is needed is a recapture protocol that removes the heating altogether, at a minimal price in duration.

The key ingredient was recently supplied in a different context. For time-shared ``blinking'' tweezers, Kim, Kim, and Ahn derived, and demonstrated experimentally, the condition under which one free--trap--free cycle restores the phase-space distribution of a trapped atom~\cite{KimAhn2025}. Here we develop that refocusing condition, obtained in Ref.~\cite{KimAhn2025} at the level of classical distributions, into a complete, quantum-mechanically exact recapture tool for gate cycles. Section~\ref{sec:rr} sets up the starting point, the exact scaling (Ermakov) solution for a wave packet released from the trap~\cite{Lewis1969,LewisLeach1982,DharaLawande1984,Kagan1996,CastinDum1996}. Section~\ref{sec:echo} develops the theory of the echo cycle in a harmonic trap, showing that, with a specific timing and using only intensity switching of the existing trap beam, the atom is recaptured exactly as if it had been held in the unperturbed trap; populations, coherences, and thermal states are all returned without heating. The timing generalizes in closed form to an arbitrary catch depth, at which the second dark window remains equal to the gate duration $T$. We show that the delta-kick ``matter-wave echo''~\cite{Chu1986,Ammann1997,Morinaga1999,Kovachy2015,Dupays2021} is the infinite-depth limit of our scheme. The two-switch sequence is further proved to be the unique time-optimal recovery throughout the experimentally relevant regime (Appendix~\ref{app:pmp}). Section~\ref{sec:numerics} validates the protocol with wave-packet simulations, including a realistic anharmonic Gaussian tweezer, finite switching ramps, and control errors. Section~\ref{sec:gates} then quantifies the impact on a realistic cesium Rydberg-gate cycle. Motivated by the unmatched axial residual in that example, Sec.~\ref{sec:multimode} develops exact refocusing of several nondegenerate harmonic modes with one global intensity waveform: we prove local existence of smooth radial--axial solution families, map their continuation over the gate-relevant parameter region, construct an exact fully anisotropic three-frequency echo which can be applied to a 3D harmonic trap, and establish an exact algebraic shortest-duration result within the minimal two-pulse palindromic family. Section~\ref{sec:anharm} extends the control beyond the harmonic approximation with a short composite sequence that cancels the leading quartic heating. Section~\ref{sec:discussion} discusses the relation to shortcuts to adiabaticity (STA)~\cite{Chen2010,Schaff2010,Schaff2011,Stefanatos2010,GueryOdelin2019}, experimental requirements, and extensions. Finally, we discuss how the scheme enables quantum computation with neutral atoms held in a single shared trapping field, such as an optical lattice.

\section{Exact dynamics of a released wave packet}
\label{sec:rr}

In this section we model the tweezer as a three-dimensional harmonic potential with frequencies $\omega_i$ ($i=x,y,z$). Anharmonic corrections are quantified numerically in Sec.~\ref{sec:numerics}, and composite echo pulses for anharmonic traps are discussed in Sec.~\ref{sec:anharm}. Unless stated otherwise we use natural units $\hbar=m=\omega=1$ for each axis. The building block for everything that follows is the exact evolution of the trapped wave packet when the trap intensity is varied, and in particular, switched off.

The key tool is the scale invariance of quadratic Hamiltonians~\cite{Lewis1969,LewisLeach1982,DharaLawande1984}: for $H(t)=p^2/2+\omega^2(t)x^2/2$, every solution $\phi(x,t)$ of the \emph{static} problem with frequency $\omega_0$ generates a solution of the time-dependent problem,
\begin{equation}
\psi(x,t)=\frac{e^{i\dot b\,x^2/2b}}{\sqrt{b(t)}}\,
\phi\!\left(\frac{x}{b(t)},\,\tau(t)\right),
\quad
\tau(t)\equiv\!\int_0^t\!\frac{\dd t'}{b^2(t')},
\label{eq:scaling}
\end{equation}
provided the \emph{scale factor} $b(t)$, which sets the instantaneous width of the wave packet in units of its initial width, obeys the Ermakov equation
\begin{equation}
\ddot b+\omega^2(t)\,b=\frac{\omega_0^2}{b^3},
\qquad b(0)=1,\ \dot b(0)=0 .
\label{eq:ermakov}
\end{equation}
For a sudden release, $\omega(t>0)=0$, the solution is $b(t)=\sqrt{1+t^2}$, and the $n$-th eigenstate expands self-similarly,
\begin{align}
\psi_n(t,x)=\;&\frac{ e^{-i\left[(n+\frac12)\arctan t-\frac{x^2 t}{2(1+t^2)}\right]} }
{\sqrt{2^n n!}\,\pi^{1/4}(1+t^2)^{1/4}}\,
e^{-\frac{x^2}{2(1+t^2)}}\, H_n\!\Big(\tfrac{x}{\sqrt{1+t^2}}\Big),
\label{eq:psin}
\end{align}
with $H_n$ the Hermite polynomials: every eigenstate expands self-similarly, and the chirp phase $\propto x^2 t/(1+t^2)$ records the position--momentum correlation that will be central to the protocol of Sec.~\ref{sec:echo}.

We note that the same scaling solution is the basis of quantum treatments of release-and-recapture (R\&R) thermometry, the standard survival diagnostic for single tweezer-trapped atoms~\cite{Tuchendler2008}. Near the motional ground state the classical Monte Carlo analysis of the R\&R signal systematically overestimates the temperature, because it carries no zero-point motion; quantum calculations of the recapture probability (free expansion of trap eigenstates followed by projection onto the bound states) correct the bias~\cite{Hoelzl2023,Biagioni2025}. A closed-form analysis of quantum recapture within the present framework, including the statistics of several identical atoms, is presented in Appendix~\ref{app:rr}.

\section{The harmonic echo protocol}
\label{sec:echo}

\subsection{Phase-plane formulation and the cost of a sudden catch}

This section develops the theory in one dimension (one motional mode), while the generalization to a realistic three-dimensional trap, with a single intensity control acting on all modes at once, is taken up in Sec.~\ref{sec:multimode}. Throughout, the classical phase-space refocusing condition of Ref.~\cite{KimAhn2025} will be promoted to an \emph{exact quantum statement}: the recaptured wave function is shown (Sec.~\ref{subsec:pure}) to equal the initial one evolved under the static trap, so that populations \emph{and} coherences are preserved. We assume a mandatory dark window of duration $T$ (e.g., the two-qubit gate), and seek the shortest trap-intensity program $u(t)\equiv\omega^2(t)/\omega_0^2$ that returns the atom \emph{exactly} to its initial motional state. The control $u(t)$ is a \emph{continuous} variable, bounded by the available laser power, $u\in[0,u_{\max}]$; we write $u_{\max}=\Lambda^2$, so $\Lambda$ is the maximum trap frequency in units of the nominal one. The most important case is $\Lambda=1$ (catch at the nominal depth). We show in Sec.~\ref{sec:optimal} that the time-optimal solution is \emph{bang-bang}, using only the endpoints $u=0$ and $u=\Lambda^2$; the intermediate segments below are anticipations of that result.

All the information required is carried by the single scalar $b(t)$ of Eq.~\eqref{eq:ermakov} ($b=1$ for an atom at rest in the nominal trap). In natural units of the nominal trap ($\omega_0=1$), and in terms of the control $u(t)$, Eq.~\eqref{eq:ermakov} is the equation of motion for the whole analysis that follows,
\begin{equation}
\ddot b=-u(t)\,b+\frac{1}{b^3}.
\label{eq:ermb}
\end{equation}
By Eq.~\eqref{eq:scaling}, the atom is back in its initial state (up to the residual trap phase evolution, made precise in Sec.~\ref{subsec:pure}) if and only if $(b,\dot b)=(1,0)$, for \emph{any} initial state. During free flight ($u=0$) the quantity
\begin{equation}
\varepsilon_0=\tfrac12\dot b^2+\frac{1}{2b^2}
\label{eq:eps0}
\end{equation}
is conserved. Release from equilibrium fixes $\varepsilon_0=\tfrac12$, and at the end of the gate window the system sits at
\begin{equation}
\big(b_T,\dot b_T\big)=\Big(\sqrt{1+T^2},\ \tfrac{T}{\sqrt{1+T^2}}\Big).
\label{eq:release}
\end{equation}

The mean energy of a state that was initially thermal with occupation $\bar n$ is $\langle H\rangle=(\bar n+\tfrac12)\hbar\omega\,(\dot b^2+b^2+b^{-2})/2$. A \emph{sudden} recapture at the nominal depth therefore heats by
\begin{equation}
\Delta\bar n_{\rm sud}=\big(\bar n+\tfrac12\big)\frac{(\omega T)^2}{2},
\label{eq:naive}
\end{equation}
i.e., $(\omega T)^2/4$ quanta for a ground-state atom. Throughout, heating is quoted \emph{per motional mode} unless a sum over modes is stated explicitly.

It is tempting to instead catch at a ``matched'' depth chosen so that the expanded width fits the reopened trap. This barely helps: the expanded state is a \emph{sheared} Gaussian, $\sigma_x\sigma_p=\tfrac12\sqrt{1+T^2}>\tfrac12$ due to the position--momentum correlation. For an initially ground-state atom, optimizing the catch frequency yields at best
\begin{equation}
\bar n_{\rm min}=\tfrac12\big(\sqrt{1+T^2}-1\big)
=\frac{T^2}{4}-\frac{T^4}{16}+\dots,
\label{eq:matched}
\end{equation}
identical to the sudden catch at leading order. The heating is caused by the correlation, not by the width mismatch, and no static trap can remove a correlation instantaneously. What is needed is a protocol that \emph{reverses} it.

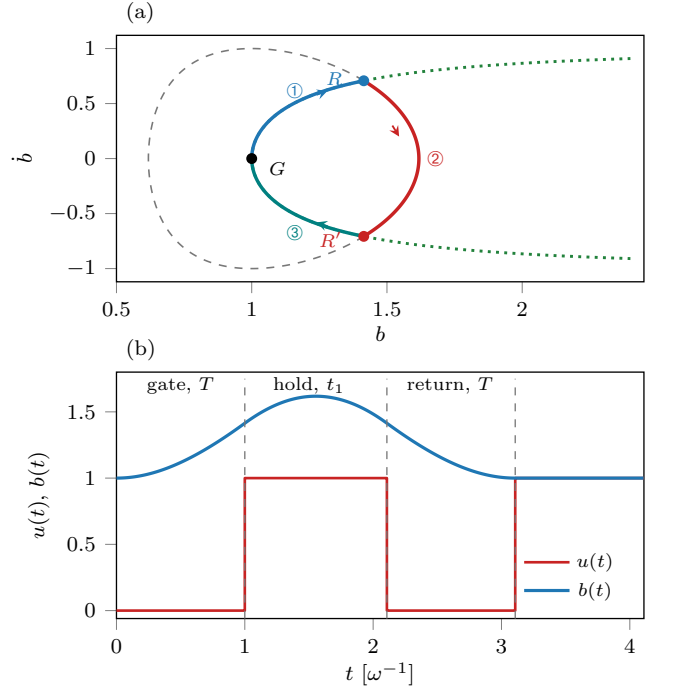
\begin{figure}[tb]
\centering
\begin{tikzpicture}
	\begin{groupplot}[group style={group size=1 by 2, vertical sep=34pt},
		rrfig, height=0.56\columnwidth, clip mode=individual,
		xlabel style={yshift=3pt}]
		\nextgroupplot[xlabel={$b$}, ylabel={$\dot b$},
		xmin=0.5,xmax=2.45,ymin=-1.12,ymax=1.12]
		\addplot[cGray, line width=0.6pt, dashed] table[x=b,y=bd]{figs/orbit.dat};
		\addplot[cGreen, dotted, line width=1.1pt] table[x=b,y=bdp]{figs/coast.dat};
		\addplot[cGreen, dotted, line width=1.1pt] table[x=b,y=bdm]{figs/coast.dat};
		\addplot[cBlue, line width=1.3pt] table[x=b,y=bd]{figs/traj_seg1.dat};
		\addplot[cRed,  line width=1.3pt] table[x=b,y=bd]{figs/traj_seg2.dat};
		\addplot[cTeal, line width=1.3pt] table[x=b,y=bd]{figs/traj_seg3.dat};
		\addplot[only marks, mark=*, mark size=1.6pt, black] coordinates {(1,0)};
		\addplot[only marks, mark=*, mark size=1.6pt, cBlue] coordinates {(1.41421,0.70711)};
		\addplot[only marks, mark=*, mark size=1.6pt, cRed] coordinates {(1.41421,-0.70711)};
		\node[font=\scriptsize, anchor=west] at (axis cs:1.03,-0.09) {$G$};
		\node[font=\scriptsize, cBlue, anchor=east] at (axis cs:1.37,0.73) {$R$};
		\node[font=\scriptsize, cRed, anchor=east] at (axis cs:1.37,-0.73) {$R'$};
		\node[font=\footnotesize, cBlue] at (axis cs:1.16,0.62) {\ding{192}};
		\node[font=\footnotesize, cRed] at (axis cs:1.68,0.0) {\ding{193}};
		\node[font=\footnotesize, cTeal] at (axis cs:1.16,-0.67) {\ding{194}};
		\draw[-stealth, cBlue, line width=0.7pt] (axis cs:1.24,0.58) -- (axis cs:1.28,0.615);
		\draw[-stealth, cRed, line width=0.7pt] (axis cs:1.52,0.30) -- (axis cs:1.545,0.20);
		\draw[-stealth, cTeal, line width=0.7pt] (axis cs:1.28,-0.615) -- (axis cs:1.24,-0.58);
		\node[anchor=south west, font=\footnotesize, yshift=1pt] at (rel axis cs:0,1) {(a)};
		\nextgroupplot[xlabel={$t\ [\omega^{-1}]$}, ylabel={$u(t)$, $b(t)$},
		xmin=0,xmax=4.11,ymin=-0.06,ymax=1.8,
		legend pos=south east]
		\addplot[cRed, line width=1.0pt] table[x=t,y=u]{figs/timeline.dat};
		\addlegendentry{$u(t)$}
		\addplot[cBlue, line width=1.2pt] table[x=t,y=b]{figs/timeline.dat};
		\addlegendentry{$b(t)$}
		\draw[cGray, dashed, line width=0.5pt] (axis cs:1,-0.06) -- (axis cs:1,1.75);
		\draw[cGray, dashed, line width=0.5pt] (axis cs:2.1071,-0.06) -- (axis cs:2.1071,1.75);
		\draw[cGray, dashed, line width=0.5pt] (axis cs:3.1071,-0.06) -- (axis cs:3.1071,1.75);
		\node[font=\scriptsize] at (axis cs:0.5,1.7) {gate, $T$};
		\node[font=\scriptsize] at (axis cs:1.5,1.7) {hold, $t_1$};
		\node[font=\scriptsize] at (axis cs:2.6,1.7) {return, $T$};
		\node[anchor=south west, font=\footnotesize, yshift=1pt] at (rel axis cs:0,1) {(b)};
	\end{groupplot}
\end{tikzpicture}
\caption{\textbf{The echo protocol in the Ermakov phase plane} (shown for $\omega T=1$).
	(a) The free-flight invariant curve $\varepsilon_0=\tfrac12$ (green dotted) passes through both the target $G=(1,0)$ and the release point $R$, Eq.~\eqref{eq:release}, on opposite (expanding/contracting) branches. With the trap on at nominal depth, $G$ is the center of the phase portrait and the closed orbit through $R$ (gray dashed) re-crosses the invariant curve exactly at the conjugate point $R'=(b_T,-\dot b_T)$. The protocol therefore consists of the gate flight \ding{192}, a hold \ding{193} of duration $t_1$ [Eq.~\eqref{eq:t1}] that transfers $R\to R'$, and a return flight \ding{194} of duration exactly $T$ that coasts into $G$, where the trap is switched on permanently.
	(b) Control $u(t)$ and scale factor $b(t)$ for the same cycle; $b$ returns to $1$ with $\dot b=0$ and remains stationary thereafter. Here $t_1=\arctan(2/\omega T)\approx1.11\,\omega^{-1}$; at higher catch depth the hold shortens according to Eq.~\eqref{eq:t1gen}, while the return window remains exactly $T$ at every depth.}
\label{fig:portrait}
\end{figure}

\subsection{Two switches suffice: the breathing-mode echo}

The reversal follows from a simple geometric fact, illustrated in Fig.~\ref{fig:portrait}(a). The target $G=(1,0)$ also lies on the invariant curve $\varepsilon_0=\tfrac12$: it is the point from which the original expansion departed. The release point after the gate, $R=(b_T,\dot b_T)$, and its conjugate, $R'=(b_T,-\dot b_T)$, sit on the same free-flight orbit, merely on opposite branches ($\dot b>0$ expanding versus $\dot b<0$ contracting) extending from $G$. All that is required for the system to return to $G$ is a maneuver that carries the system from the expanding branch to the contracting one: a chirp reversal of the breathing mode, in direct analogy to the $\pi$ pulse of a spin echo.

The nominal trap performs exactly this maneuver. With $u=\Lambda^2$ held constant, $\varepsilon_\Lambda=\tfrac12\dot b^2+\tfrac12\Lambda^2b^2+\tfrac{1}{2b^2}$ is conserved, and since $\varepsilon_\Lambda=\varepsilon_0+\tfrac12\Lambda^2 b^2$, the trajectory returns to the curve $\varepsilon_0=\tfrac12$ precisely when $b$ returns to $b_T$, now with $\dot b=-\dot b_T$. Two consequences follow immediately. First, the duration of the second dark window is
\begin{equation}
t_2=T \qquad\text{for every catch depth }\Lambda,
\label{eq:t2}
\end{equation}
because the return flight from $(b_T,-\dot b_T)$ is the exact time reverse of the original expansion. Second, the hold time is the arc time from $(b_T,+\dot b_T)$ to $(b_T,-\dot b_T)$ along the closed orbit, which evaluates in closed form (Appendix~\ref{app:tau}) to
\begin{align}\label{eq:t1gen}
&t_1(\Lambda,T)=\\
&\frac{1}{\Lambda}\left[\frac{\pi}{2}-
\arcsin\frac{\Lambda^2(1+T^2)-1}
{\sqrt{\big[(\Lambda\!-\!1)^2\!+\!\Lambda^2T^2\big]\big[(\Lambda\!+\!1)^2\!+\!\Lambda^2T^2\big]}}\right].\nonumber
\end{align}
For the practically central case $\Lambda=1$ this reduces to
\begin{equation}
	\ t_1=\frac{\pi}{2}-\arctan\frac{T}{2}\ 
	\label{eq:t1}
\end{equation}
in units of $\omega^{-1}$. An equivalent form of Eq.~\eqref{eq:t1gen} is $t_1=(\Lambda\omega)^{-1}\arctan\!\big[2\Lambda\omega T/\big(\Lambda^2(1+\omega^2T^2)-1\big)\big]$, whose denominator is strictly positive for $T>0$ and $\Lambda\ge1$. The complete cycle is therefore
\[
\text{off}(T)\ \to\ \text{on}(t_1)\ \to\ \text{off}(T)\ \to\ \text{on},
\]
with a total duration of $T_{\rm cyc}=2T+t_1$ and post-gate overhead $t_1+T$. 

In the limit $T\to\infty$, $t_1\to 2/(\Lambda^2 T)$: the hold degenerates into an impulsive lens of area $\int \omega^2\dd t = 2/T$, i.e., a thin lens of focal time $T/2$, and the protocol becomes the known delta-kick ``matter-wave echo'' (a 4$f$ imaging system in time)~\cite{Chu1986,Ammann1997,Morinaga1999,Kovachy2015,Dupays2021}. Equation~\eqref{eq:t1gen} completes this picture at finite strength: it gives the exact hold time at every finite intensity, with the delta kick as its $\Lambda\rightarrow\infty$ limit, so exact refocusing never requires intensity beyond the nominal depth, only a longer hold. At the nominal depth, this timing coincides with the single-cycle refocusing condition obtained for blinking tweezers in Ref.~\cite{KimAhn2025}, there derived at the level of the classical phase-space distribution and confirmed experimentally. The remainder of this section and Appendices~\ref{app:tau} and \ref{app:pmp} establish the quantum content, finite-depth generality, and optimality of the cycle.

\subsection{The cycle is pure time evolution}
\label{subsec:pure}

The protocol ends with $(b,\dot b)=(1,0)$, so by Eq.~\eqref{eq:scaling} the final wave function is $\psi(x,T_{\rm cyc})=\phi(x,\tau)$, where $\phi$ evolves under the \emph{static} trap Hamiltonian $H_0$. The entire cycle is thus the unitary
\begin{equation}
U_{\rm cyc}=\exp(-iH_0\tau/\hbar),
\qquad
\tau=2\arctan T+\tau_1,
\label{eq:Ucyc}
\end{equation}
with $\tau_1<t_1$ given in closed form in Appendix~\ref{app:tau}. This statement is far stronger than the return of a particular state: the cycle preserves arbitrary motional \emph{coherences}, because it is operationally indistinguishable from holding the atom in the unperturbed trap for a time $\tau$. Concretely, a coherent superposition of motional states is returned with only the trap phases $e^{-iE_n\tau/\hbar}$ between its components; as special cases, every trap eigenstate returns to itself and a thermal state of any temperature is returned unchanged. 

The protocol is state independent, requires no knowledge of the atomic temperature, and, since only the intensity is switched, requires no optical phase control of the trap light. A phase-space displacement already present before the complete cycle is rotated by the angle $\tau$ with no change of oscillator energy. A displacement imparted \emph{during} the cycle is acted on by the remaining partial propagator and is treated in Sec.~\ref{sec:gates}. The echo removes breathing (width) excitation exactly but does not remove coherent displacements, which must be budgeted separately~\cite{Robicheaux2021}.

Gravity is negligible on gate time scales. During a typical $T=1\,\mu$s, an atom falls $gT^2/2\approx5\,$pm and acquires a velocity $gT$, both of which are $\lesssim10^{-3}$ of the ground-state position and velocity widths for the parameters of Sec.~\ref{sec:gates}. 

\subsection{Time optimality}
\label{sec:optimal}

A natural question is whether an even faster heating-free cycle is possible. For bounded control $u\in[0,1]$ the answer is provably no throughout the experimentally relevant regime. In Appendix~\ref{app:pmp} we analyze the minimum-time recovery problem with Pontryagin's maximum principle. Since no singular arcs exist, every optimal control is bang-bang. Switches from free flight to the trap can occur only while the packet expands, and the reverse switches only while it contracts. Hence, the final arc is always free flight. The intuition is simple: a switch into the trap is only useful while the packet is expanding, and a switch back to free flight only while it is contracting, so the trap is used exactly once, for the single hold that reverses the breathing chirp. We show that the unique single-hold extremal (i.e., the two-switch protocol of Sec.~\ref{sec:echo}) is exactly Eqs.~\eqref{eq:t1} and \eqref{eq:t2}, and every extremal with an extra pair of switches is strictly slower because each interior free/trap pair costs more than $\pi$ in time (Appendix~\ref{app:pmp}). Consecutive interior arcs are shown to last $1/r$ (free) and $\pi/2+\arctan r$ (trapped), with $r=|\dot b|/b$ at the switches, which gives a lower bound of $4.836\,\omega^{-1}$ for any extremal with three or more switches. The result is a theorem that for all $0<\omega T\le 4.4107$ the two-switch protocol is the \emph{unique} globally time-optimal recovery among all measurable intensity programs $0\le u(t)\le1$, within the harmonic, instantaneous-switching control model. 

The restriction is essential, since for long dark windows the claim is incorrect. For example, at $\omega T=10$ an explicit four-switch extremal that transiently compresses the packet to $b\approx0.32$ recovers in $8.17\,\omega^{-1}$ instead of $10.20\,\omega^{-1}$. This family of pulses first overtakes the two-switch protocol near $\omega T\approx6.90$ (Appendix~\ref{app:pmp}). Importantly, all gate-relevant windows ($\omega T\lesssim1$) lie deep inside the certified regime, and the direct numerical search over four-segment bang-bang sequences (Sec.~\ref{sec:numerics}) is consistent with the theorem. 

If intensity headroom is available ($\Lambda>1$), Eq.~\eqref{eq:t1gen} shortens the hold, with $t_1\to 2T/[\Lambda^2(1+T^2)]$ as $\Lambda\to\infty$, at the price of $\Lambda^2$ times the nominal power. Within the single-hold family the return window equals $T$ at every depth [Eq.~\eqref{eq:t2}], and the theorem generalizes: for $0\le u\le\Lambda^2$ the single-hold sequence remains the unique global optimum whenever $\omega T$ lies below a certified bound $x_{\rm cert}(\Lambda)$ derived in Appendix~\ref{app:pmp}, with $x_{\rm cert}(2)=0.923$ comfortably covering the operating point of Sec.~\ref{sec:gates} at fourfold power. The $T$ floor itself, however, is not optimal at large headroom: beyond $\Lambda\approx5.6$ at that operating point, sequences that transiently overcompress the packet recover in less than $T$ (Appendix~\ref{app:pmp}).

\subsection{Using the two dark windows: gate placement and split gates}
\label{subsec:windows}

The cycle offers two equal dark windows, and either or both can host the gate. Since off$(T)$--on$(t_1)$--off$(T)$ is time symmetric, the gate may equally well be executed in the \emph{second} window, which a single switch-on then closes with zero post-gate dark time (up to one switching ramp). The wave packet is there contracting rather than expanding, with the same variances and the opposite position--momentum correlation. This ordering is preferable when the recovery time is cheaper before the gate than after it.

The windows can also be used together. Executing a blockade gate as two $\sqrt{\rm CZ}$ halves with the Doppler detuning reversed in between is an established way to cancel the error from a quasi-static atomic velocity, the reversal being obtained either by reversing the beam direction or by the atoms' own harmonic motion~\cite{Jandura2023}. Dressing-based schemes offer a complementary route~\cite{LiQian2025}. The echo cycle supplies precisely the structure this requires: two equal free-flight windows separated by a trap interlude. Below, we evaluate the ability of gate splitting within our echo sequence to mitigate the two central gate errors associated with velocity.

\emph{Doppler error.} The interlude is the hold, which rotates each atom's phase-space coordinates by the angle $\omega t_1$ rather than by half a trap period, so the velocity echo is partial. The velocity is constant within each window; across the hold it maps to $v_2=-\omega x_1\sin\omega t_1+v_1\cos\omega t_1$, so a thermal ensemble carries the correlation $\langle v_1v_2\rangle=-\sigma_v^2\cos\omega t_1$. At the level of the Doppler-phase proxy of Sec.~\ref{sec:gates}, splitting the interrogation equally between the windows (equal weights are optimal) multiplies the Doppler error variance by $(1\mp\cos\omega t_1)/2$ relative to the unsplit gate, where the upper sign applies to halves driven by the \emph{same} beams and the lower to a reversed effective wave vector. This means that the hold's partial velocity reversal makes the same-beam configuration the echoed one, so the suppression requires no beam-reversal hardware. With $\cos\omega t_1=\omega T/\sqrt{4+\omega^2T^2}=0.27$ at the operating point of Sec.~\ref{sec:gates}, the same-beam split reduces the Doppler error variance by a factor $2.7$, taking $\epsilon_D$ of Table~\ref{tab:cs} from $4.2\times10^{-4}$ to $1.5\times10^{-4}$. The improvement factor improves with $T$, recovering the complete cancellation of Ref.~\cite{Jandura2023} in the delta-kick limit $T\to\infty$, where the hold reverses the velocity exactly.

\emph{Photon recoil.} Recoil behaves differently, because a momentum kick is transformed by the remainder of the cycle from the instant it is imparted. Reduce each half to a single net two-photon impulse of magnitude $\hbar k_{\rm eff}$, imparted at an adjustable time within its window and with either sign. Exact closure of the two kicks in phase space is then impossible. This is because cancellation of the momentum components requires $|\cos\omega t_1-\omega a\sin\omega t_1|=1$, with $a\in[0,T]$ the flight time remaining after the first kick, whereas $|\cos\omega t_1-\omega a\sin\omega t_1|\le\cos\omega t_1<1$ over the whole window. Optimal placement still helps: kicks of opposite sign at the end of the first window and the start of the second reduce the residual displacement energy by a factor $4.7$ relative to coincident kicks, leaving a floor of $0.42\,(k_{\rm eff}a_{\rm ho})^2$ quanta per branch, which is about $0.02$ quanta for the parameters of Table~\ref{tab:cs}, comparable to a single unclosed kick. 

The obstruction, however, is specific to the single-impulse reduction. Shifting one impulse inside the first window and one inside the second moves the final displacement along two linearly independent phase-space directions, so a pulse train with two or more impulses per window can, in principle, provide enough timing freedom to satisfy the two closure conditions. Real $\sqrt{\rm CZ}$ pulse trains contain several impulses, and their timings within a window are free at fixed pulse areas because the velocity is constant during free flight; the same constancy leaves the conditional phases unaffected by such timing shifts. Recoil closure can therefore be treated as a pulse-design constraint.

\section{Numerical validation and realistic conditions}
\label{sec:numerics}

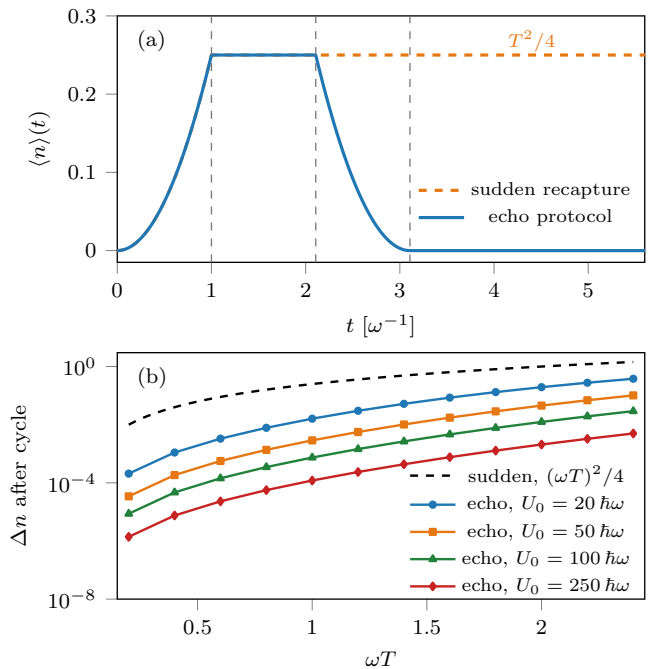
\begin{figure}[tb]
\centering
\begin{tikzpicture}
\begin{groupplot}[group style={group size=1 by 2, vertical sep=34pt},
  rrfig, height=0.56\columnwidth]
\nextgroupplot[xlabel={$t\ [\omega^{-1}]$}, ylabel={$\langle n\rangle(t)$},
  xmin=0,xmax=5.6,ymin=-0.015,ymax=0.30,
  legend style={at={(1,0.1)}, anchor=south east}]
\addplot[cOrange, line width=1.2pt, dashed] table[x=t,y=n]{figs/demo_naive.dat};
\addlegendentry{sudden recapture}
\addplot[cBlue, line width=1.2pt] table[x=t,y=n]{figs/demo_echo.dat};
\addlegendentry{echo protocol}
\draw[cGray, dashed, line width=0.5pt] (axis cs:1,-0.015) -- (axis cs:1,0.30);
\draw[cGray, dashed, line width=0.5pt] (axis cs:2.1071,-0.015) -- (axis cs:2.1071,0.30);
\draw[cGray, dashed, line width=0.5pt] (axis cs:3.1071,-0.015) -- (axis cs:3.1071,0.30);
\node[font=\scriptsize, cOrange] at (axis cs:4.4,0.268) {$T^2/4$};
\node[anchor=north west,font=\footnotesize] at (rel axis cs:0.02,0.98) {(a)};
\nextgroupplot[xlabel={$\omega T$}, ylabel={$\Delta n$ after cycle},
  xmin=0.15,xmax=2.45, ymode=log, ymin=1e-8, ymax=3,
  legend style={at={(1,-0.03)},anchor=south east}, legend columns=1]
\addplot[black, dashed, domain=0.2:2.4, samples=60]{x^2/4};
\addlegendentry{sudden, $(\omega T)^2/4$}
\addplot[cBlue, mark=*, mark size=1.1pt] table[x=T,y=dn_echo]{figs/anharm_U20.dat};
\addlegendentry{echo, $U_0=20\,\hbar\omega$}
\addplot[cOrange, mark=square*, mark size=1.1pt] table[x=T,y=dn_echo]{figs/anharm_U50.dat};
\addlegendentry{echo, $U_0=50\,\hbar\omega$}
\addplot[cGreen, mark=triangle*, mark size=1.4pt] table[x=T,y=dn_echo]{figs/anharm_U100.dat};
\addlegendentry{echo, $U_0=100\,\hbar\omega$}
\addplot[cRed, mark=diamond*, mark size=1.4pt] table[x=T,y=dn_echo]{figs/anharm_U250.dat};
\addlegendentry{echo, $U_0=250\,\hbar\omega$}
\node[anchor=north west,font=\footnotesize] at (rel axis cs:0.02,0.98) {(b)};
\end{groupplot}
\end{tikzpicture}
\caption{\textbf{Wave-packet numerical validation.}
(a) Motional excitation $\langle n\rangle(t)=\langle H_0\rangle/\hbar\omega-\tfrac12$ of an initially ground-state atom during the cycle, from split-step integration of the Schr\"odinger equation in a harmonic trap ($\omega T=1$; switch times marked). The sudden recapture retains $\Delta n=T^2/4$ exactly; the echo protocol returns $\Delta n=0$ at numerical precision ($<10^{-9}$). (b) Residual excitation after the cycle in a \emph{Gaussian} tweezer of depth $U_0$ and curvature matched to $\omega=1$, for an atom initially in the ground state of the Gaussian trap (obtained by imaginary-time propagation), using the harmonic timing of Eq.~\eqref{eq:t1}, versus the dark time $T$. The residual scales as $U_0^{-2}$ and lies orders of magnitude below the sudden-recapture heating (dashed). At a realistic depth $U_0=250\,\hbar\omega$ and $\omega T=0.55$ it is $\approx2\times10^{-5}$ quanta.}
\label{fig:demo}
\end{figure}

We validated the protocol with direct split-step integration of the time-dependent Schr\"odinger equation (Appendix~\ref{app:methods}). In a harmonic trap the numerics confirm the analytical predictions at machine precision [Fig.~\ref{fig:demo}(a)]: for $\omega T=1$ the sudden recapture leaves exactly $\Delta n=T^2/4=0.25$ quanta, while the echo returns the ground state with $\Delta n<10^{-9}$ and unit fidelity. To test Eq.~\eqref{eq:Ucyc}, we propagated the superposition $(\phi_0+\phi_2)/\sqrt2$ through the cycle and compared it with the same state evolved for the time $\tau$ of Eq.~\eqref{eq:Ucyc} under the static trap. The squared overlap of the two wave functions deviates from unity by less than $10^{-8}$, confirming that the cycle is pure time evolution, including its effect on motional coherences. An independent Ermakov-level check confirms Eqs.~\eqref{eq:t2} and \eqref{eq:t1gen} for $\Lambda=1$ and $\Lambda=2$ to $10^{-9}$, and a numerical minimum-time search over four-segment on/off sequences (sequential quadratic programming from 300 random initial duration vectors, with the refocusing endpoint imposed as a constraint; see Appendix~\ref{app:methods}) reproduced the two-switch total $T+t_1$ at the solver tolerance without finding any faster protocol, both at $\omega T=1$ and at the operating point of Sec.~\ref{sec:gates}, consistent with the optimality theorem.

\emph{Anharmonicity.} A real tweezer is Gaussian, $V(x)=-U_0e^{-2x^2/w^2}$ with $w=2\sqrt{U_0}\,a_{\rm ho}$ for a curvature-matched depth $U_0$ (in units of $\hbar\omega$). Using the \emph{harmonic} timing Eq.~\eqref{eq:t1} without any adjustment, the residual excitation after the cycle [Fig.~\ref{fig:demo}(b)] scales as $U_0^{-2}$. This is because the curvature-matched trap expands as $V+U_0=\tfrac12x^2-x^4/(8U_0)+x^6/(48U_0^2)-\dots$ (lengths in $a_{\rm ho}$), and since the echo treats the quadratic part exactly, the transition \emph{amplitude} from a trap eigenstate to any other over the cycle is first order in the quartic coefficient, $\propto U_0^{-1}$. The transferred population, which quantifies the heating, is its square, $\propto U_0^{-2}$. The fitted slope over $U_0=30$--$200\,\hbar\omega$ is $-1.98$, and the residual remains far below the sudden-catch heating over the whole practical range. At a typical depth $U_0=250\,\hbar\omega$ and $\omega T=0.55$ the residual is $\Delta n\approx2\times10^{-5}$ quanta, with no unbound population resolved above the numerical floor of $10^{-12}$. Even at a shallow trap, $U_0=20\,\hbar\omega$, and $\omega T=1$ the echo leaves $0.016$ quanta versus $0.23$ for the sudden catch. For shallow traps and long flights the harmonic $t_1$ is no longer exactly optimal. However, a one-parameter re-tune of the hold recovers most of the loss. For example, at $U_0=50$, $\omega T=1.6$, lengthening $t_1$ by $6\%$ reduces the residual from $0.017$ to $0.008$ quanta. The pair $(t_1,t_2)$ can be calibrated \emph{in situ} by minimizing measured heating.

\emph{Control imperfections.} Finite switching ramps are not intrinsically harmful. We consider linear ramps of duration $t_r$, arranged causally: the trap is first ramped from $u=1$ to $0$, the mandatory dark interval $T$ begins only once this ramp is complete, and each turn-on ramp starts only after the corresponding dark interval has ended. Within the harmonic model, re-optimizing the hold plateau restores \emph{exact} harmonic refocusing, with two end-point conditions met by two retimed durations, for every ramp length tested (with endpoint residuals remaining at solver precision up to $t_r=0.3\,\omega^{-1}\approx0.5\,\mu$s), with the second dark window remaining exactly $T$. The finite ramps simply modify the effective symplectic transformation of the hold segment without changing the number of adjustable timing parameters, so the two endpoint conditions remain exactly solvable. In the Gaussian trap of Sec.~\ref{sec:gates}, the retimed schedule keeps the residual at the anharmonic floor: $\Delta n=2.3\times10^{-5}$, $3.0\times10^{-5}$, and $6.6\times10^{-5}$ at $t_r=0.05$, $0.1$, and $0.3\,\omega^{-1}$. In this ideal linear-ramp harmonic model the motional effect of the ramps is therefore removed entirely by calibration (the four ramps still add $4t_r$ of wall-clock time, and scattering or hardware transfer-function errors are budgeted separately). For the trapezoidal ramps used here, finite switching acts primarily as an effective increase of the optimal hold duration by approximately $1.5\,t_r$, so an un-retimed sequence follows the same quadratic timing sensitivity discussed below.

Static intensity miscalibration $u=1+\epsilon$ during the hold gives $\Delta n\simeq0.86\,\epsilon^2$ ($9\times10^{-5}$ quanta at $1\%$), and hold-time jitter gives $\Delta n\simeq0.33\,(\omega\,\delta t_1)^2$ ($10^{-5}$ quanta at $10\,$ns). A radial-frequency mismatch $\delta=\Delta\omega/\omega$, from trap ellipticity $\omega_x\neq\omega_y$ or from site-to-site frequency disorder in an array, leaves the mistimed mode with $\Delta n\simeq0.80\,\delta^2$ for a ground-state atom ($8\times10^{-5}$ at $1\%$, $2\times10^{-3}$ at $5\%$), the dominant static sensitivity, consistent with the resonance splitting observed for elliptic traps in Ref.~\cite{KimAhn2025}. All coefficients are evaluated at the operating point $\omega T=0.55$ and per mode for a ground-state atom (for a thermal state they scale as $2\bar n+1$). All sensitivities are quadratic, as expected for a protocol tuned at an extremum.

The extension from one matched mode to several nondegenerate harmonic modes, including exact radial--axial and fully anisotropic common-control echoes, is developed in Sec.~\ref{sec:multimode}.

\section{Application to Rydberg-gate cycles}
\label{sec:gates}

\begin{table}[b]
\caption{Echo protocol for a cesium Rydberg-gate cycle. Trap: $U_0/k_B=1\,$mK, $w_0=0.9\,\mu$m ($U_0=237\,\hbar\omega_r$, $a_{\rm ho}=29\,$nm); dark window $T=1\,\mu$s ($\omega_r T=0.55$); initial $\bar n_r=0.1$, $\bar n_z=0.5$.}
\label{tab:cs}
\begin{ruledtabular}
\begin{tabular}{lcc}
 & sudden recapture & echo protocol \\
\hline
$\Delta \bar n_r$ per gate (two modes) & $0.15\,(2\bar n_r{+}1)$\footnotemark[1] & $0$ (exact)\footnotemark[5] \\
$\Delta \bar n_z$ per gate & $0.005\,(2\bar n_z{+}1)$ & $0.018\,(2\bar n_z{+}1)$ \\
heating-energy suppression & --- & $\times\,19$\footnotemark[4] \\
$\bar n_r$ after 20 gates & $9.8$ & $0.1$ \\
Doppler error $>1\%$ after & $\approx23$ gates & pinned at floor\footnotemark[2] \\
post-gate overhead & $0$ & $t_1{+}T=3.4\,\mu$s\footnotemark[3] \\
\end{tabular}
\end{ruledtabular}
\footnotetext[1]{Sum over both radial modes, $2\times(2\bar n+1)(\omega_rT)^2/4$, Eq.~\eqref{eq:naive}.}
\footnotetext[2]{Pinned at the $\bar n_r=0.1$ floor, $\epsilon_D\approx4.2\times10^{-4}$ for the counter-propagating geometry assumed here (Sec.~\ref{sec:gates}).}
\footnotetext[3]{$1.4\,\mu$s if the catch is performed at $\Lambda=2$ (four times the nominal power), Eq.~\eqref{eq:t1gen}; the axial residual then also drops to $0.0022(2\bar n_z+1)$ quanta per cycle.}
\footnotetext[4]{Ratio of the per-cycle heating energies summed over the three modes, $\sum_i\Delta\bar n_i\,\hbar\omega_i$, at the tabulated occupations; $\times\,31$ for a three-dimensional ground-state atom.}
\footnotetext[5]{Exact in the matched harmonic model; the one-dimensional Gaussian-tweezer residual is ${\approx}2\times10^{-5}$ quanta per radial mode at the stated depth (Sec.~\ref{sec:numerics}).}
\end{table}

\begin{figure}[tb]
\centering
\begin{tikzpicture}
\begin{groupplot}[group style={group size=1 by 2, vertical sep=34pt},
  rrfig, height=0.56\columnwidth]
\nextgroupplot[xlabel={$\omega_r T$}, ylabel={heating energy $[\hbar\omega_z]$ per cycle},
  xmin=0.08,xmax=1.52, ymode=log, ymin=3e-4, ymax=8,
  legend style={at={(1,0)},anchor=south east}]
\addplot[black, dashed] table[x=T,y=dnE_tot_naive]{figs/axial.dat};
\addlegendentry{sudden, all modes}
\addplot[cOrange, dashdotted] table[x=T,y=dnz_naive]{figs/axial.dat};
\addlegendentry{sudden, axial only}
\addplot[cBlue, line width=1.2pt] table[x=T,y=dnz_echo]{figs/axial.dat};
\addlegendentry{echo residual (axial)}
\node[anchor=north west,font=\footnotesize] at (rel axis cs:0,1) {(a)};
\nextgroupplot[xlabel={gate number}, ylabel={Doppler gate error $\epsilon_D$},
  xmin=0,xmax=40, ymode=log, ymin=1.5e-4, ymax=1.5,
  legend style={at={(0.5,0.9)},anchor=north east}]
\addplot[cOrange, dashed, line width=1.2pt] table[x=k,y=eps_naive]{figs/gates.dat};
\addlegendentry{sudden recapture}
\addplot[cBlue, line width=1.2pt] table[x=k,y=eps_echo]{figs/gates.dat};
\addlegendentry{echo protocol}
\draw[cGray, dashed, line width=0.6pt] (axis cs:0,0.01) -- (axis cs:40,0.01);
\node[font=\scriptsize, cGray, anchor=south west] at (axis cs:33,0.011) {$1\%$};
\node[anchor=north west,font=\footnotesize] at (rel axis cs:0,1) {(b)};
\end{groupplot}
\end{tikzpicture}
\caption{\textbf{Heating suppression and Doppler-error control over a circuit.}
(a) Motional heating energy per release--recapture cycle, in units of $\hbar\omega_z$, versus the dark time $T$, for a cylindrically symmetric tweezer with $\eta=\omega_r/\omega_z=3.75$ and a single global intensity control timed on the radial frequency. The echo removes the radial heating exactly in the harmonic model; the remaining axial residual (blue) lies a factor ${\approx}30$ below the total sudden-recapture heating (black dashed).
(b) Motional Doppler error proxy $\epsilon_D=\big(k_{\rm eff}\sigma_v\tau_R\big)^2/2$ versus gate number for the cesium parameters of Table~\ref{tab:cs}, assuming complete phase randomization of the breathing excitation between gates; gate number $k$ counts completed cycles, with $k=0$ the initial state. With sudden recapture the occupation grows geometrically, $\bar n_k=(\bar n_0+\tfrac12)(1+(\omega_rT)^2/2)^k-\tfrac12$, and the error crosses $1\%$ near the 23rd gate; with the echo the error remains pinned at the initial-temperature floor.}
\label{fig:gates}
\end{figure}
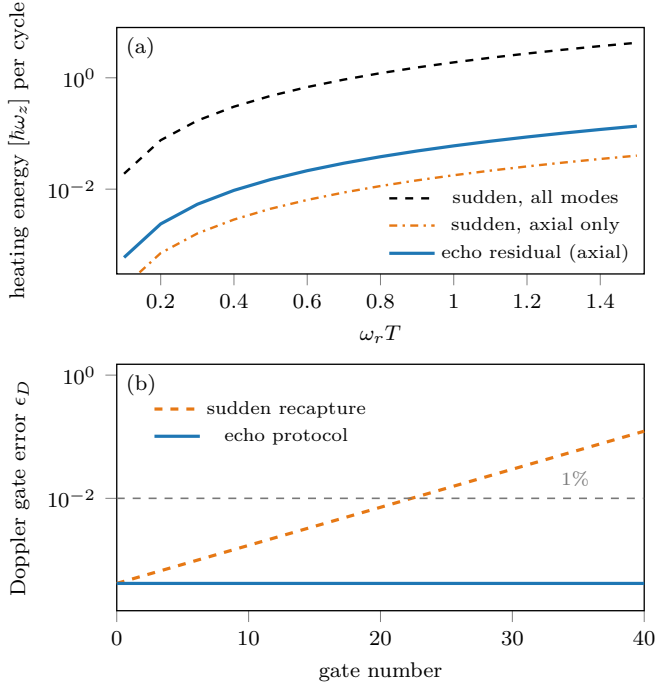

We now attach numbers relevant to a neutral-atom quantum processor, taking cesium as a concrete example. We consider a tweezer of waist $w_0=0.9\,\mu$m at wavelength $\lambda=1064\,$nm and depth $U_0/k_B=1\,$mK. These parameters give $\omega_r=2\pi\times88\,$kHz, $\omega_z=2\pi\times23.5\,$kHz ($\eta=3.75$), a ground-state size $a_{\rm ho}=29\,$nm, and $U_0=237\,\hbar\omega_r$. This is the regime of Fig.~\ref{fig:demo}(b), where the anharmonic residual of the echo is at the $10^{-5}$ level. For the dark window we take $T=1\,\mu$s, comprising the entangling pulse sequence ($\sim$0.3--0.5$\,\mu$s~\cite{Evered2023,Jandura2022}) and switching margins, so $\omega_r T=0.55$. In what follows, we assess the performance of the motional refocusing scheme in this realistic scenario, and the results are summarized in Table~\ref{tab:cs}.

With sudden recapture, Eq.~\eqref{eq:naive} gives $0.076$ quanta per radial mode per gate ($0.15$ over both radial modes) and $0.005$ axially, for a ground-state atom. Because the heating is proportional to $\bar n+\tfrac12$, the occupation grows \emph{geometrically}: assuming the breathing excitation dephases between gates (anharmonicity and intervening operations randomize its phase), $\bar n_{k+1}=\bar n_k+(\bar n_k+\tfrac12)(\omega_rT)^2/2$, giving $\bar n_r\approx9.8$ after only 20 gates and $\approx23$ after 26, at which point the mean energy of each radial mode alone reaches a tenth of the trap depth and the harmonic description itself begins to fail. Fully coherent repetition, in contrast, is governed by a product of symplectic maps and can remain bounded or become parametrically unstable, depending on the inter-gate trap phase. 

The echo protocol removes the radial channel exactly in the harmonic model, while in the Gaussian trap the anharmonic radial residual at this depth is ${\approx}2\times10^{-5}$ quanta per mode (Sec.~\ref{sec:numerics}). Therefore, the axial residual of $0.018(2\bar n_z+1)$ axial quanta per cycle dominates the heating that remains. Since a radial quantum carries $\eta=3.75$ times the energy of an axial one, the per-cycle heating energy is suppressed by $\times31$ for a ground-state atom, and the dominant radial contribution to the motional energy budget is suppressed to the anharmonic floor. When the added latency is acceptable, the common-control sequences of Sec.~\ref{sec:multimode} remove the axial residual as well and give exact harmonic closure of all three axes.

The motional temperature enters the gate error through the Doppler dephasing of the Rydberg excitation. To be concrete, we take $\epsilon_D=(k_{\rm eff}\sigma_v\tau_R)^2/2$~\cite{Saffman2010,deLeseleuc2018}, with $k_{\rm eff}=2\pi(1/459-1/1038)\,{\rm nm}^{-1}=7.6\times10^6\,$m$^{-1}$ for counter-propagating two-photon excitation of Cs, with the beams taken to propagate along a refocused radial axis so that the relevant velocity spread is radial, $\sigma_v=\sqrt{\hbar\omega_r(2\bar n_r+1)/2m}$ ($11.5\,$mm/s at $\bar n_r\!\approx\!0$), and $\tau_R=0.3\,\mu$s of Rydberg manipulation time; this evaluates to $\epsilon_D\approx4.2\times10^{-4}$ at $\bar n_r=0.1$, consistent in magnitude with published error budgets~\cite{deLeseleuc2018,Pagano2022,Evered2023}, and it scales strictly as $(2\bar n_r+1)$. For an anisotropic trap the proxy generalizes to $\sigma_\delta^2=\sum_i k_{{\rm eff},i}^2(\hbar\omega_i/2m)(2\bar n_i+1)$: excitation predominantly along the axial direction would sample the one mode the radially timed echo does not refocus, so the beam geometry matters. Figure~\ref{fig:gates}(b) shows the consequence: with sudden recapture the Doppler error crosses $1\%$ near the 23rd gate of a circuit and then runs away, whereas with the echo it remains essentially at the initial-cooling floor in this radial-beam geometry. A nonzero axial component of $\mathbf k_{\rm eff}$ would introduce slow growth through the axial residual. In circuit terms, the echo converts the radial Doppler contribution, the dominant motional error, from a depth-limiting error into a constant, calibratable offset.

The post-gate overhead is $t_1+T=2.35+1.0=3.4\,\mu$s at nominal power, or $1.4\,\mu$s if the catch is performed at twice the trap frequency ($\Lambda=2$). 
This is negligible compared with rearrangement and transport intervals ($\sim100\,\mu$s~\cite{Bluvstein2024}) and small compared with typical measurement or cooling blocks. 
For atoms near the ground state the associated leakage out of the trap is exponentially small, 
the wave packet stays deep inside the trap volume and the survival remains essentially unity. 


\section{Exact multimode harmonic refocusing}
\label{sec:multimode}

The gate example of Sec.~\ref{sec:gates} used a sequence timed to the radial frequency. It therefore refocuses both radial axes under cylindrical symmetry, but it leaves a small axial breathing excitation because $\omega_z\neq\omega_r$. We now show that the same single global intensity control can refocus several nondegenerate harmonic modes exactly, at the cost of a few additional trap-on and trap-off intervals. Common-control shortcuts for several oscillators have been developed for other boundary conditions~\cite{Palmero2015,Tobalina2020}, and frequency-jump protocols have been proposed for simultaneous three-dimensional squeezing~\cite{Marocco2026}. Here, in contrast, the boundary condition is the prescribed common dark interval of a trap-off gate, followed by exact return of every mode to its equilibrium oscillator.

The connection with the Ermakov description is direct. For each principal axis $i$, the common intensity waveform $u(t)$ drives
\begin{equation}
 \ddot b_i+\omega_i^2u(t)b_i=\frac{\omega_i^2}{b_i^3},
 \qquad b_i(0)=1,\quad \dot b_i(0)=0.
 \label{eq:mm_ermakov_main}
\end{equation}
Exact harmonic refocusing of all modes means
\begin{equation}
 b_i(t_f)=1,\qquad \dot b_i(t_f)=0
 \quad\text{for every }i.
 \label{eq:mm_endpoints_main}
\end{equation}
Geometrically, the shared intensity waveform drives a different trajectory in the Ermakov plane $(b_i,\dot b_i)$ of each mode, and the multimode problem is to close all of these trajectories at the same physical time. A single hold supplies enough timing freedom for one frequency. The additional on/off intervals below provide the extra independent controls needed to close the remaining modes.

We measure all segment durations below in units of $\omega_r^{-1}$, write $x=\omega_rT$, and use the frequency ratios
\begin{equation}
 \eta_i\equiv\frac{\omega_i}{\omega_r}.
 \label{eq:mm_eta_main}
\end{equation}
Thus $\eta_x=\eta_y=1$ in a cylindrically symmetric trap, while $\eta_z=1/\eta$ with the aspect ratio $\eta\equiv\omega_r/\omega_z$ used in Sec.~\ref{sec:gates}.

For constructing the pulse sequence it is convenient to use a matrix representation equivalent to Eq.~\eqref{eq:mm_ermakov_main}. Define the normalized operator-valued Heisenberg quadrature vector of mode $i$,
\begin{equation}
 \hat{\bm z}_i\equiv
 \begin{pmatrix}\hat Q_i\\ \hat P_i\end{pmatrix},\qquad
 \hat Q_i=\sqrt{\frac{m\omega_i}{\hbar}}\,\hat x_i,
 \quad
 \hat P_i=\frac{\hat p_i}{\sqrt{m\hbar\omega_i}}.
 \label{eq:mm_quad_main}
\end{equation}
Let $X(s)$ denote a trap-off interval of physical duration $s/\omega_r$, and let $Y(s)$ denote a trap-on interval of the same duration at frequency $\Lambda\omega_i$, corresponding to intensity $u=\Lambda^2$. Their action on $\hat{\bm z}_i$ is
\begin{align}
 F_i(s)&=\begin{pmatrix}1&\eta_i s\\0&1\end{pmatrix},\nonumber\\
 R_{i,\Lambda}(s)&=\begin{pmatrix}
 \cos(\Lambda\eta_i s)&\Lambda^{-1}\sin(\Lambda\eta_i s)\\
 -\Lambda\sin(\Lambda\eta_i s)&\cos(\Lambda\eta_i s)
 \end{pmatrix}.
 \label{eq:mm_elements_main}
\end{align}
For any complete sequence, $M_i$ denotes the ordered product of these matrices, with the earliest interval on the right, so that
\begin{equation}
 \hat{\bm z}_i(t_f)=M_i\hat{\bm z}_i(0).
 \label{eq:mm_complete_main}
\end{equation}
The endpoint conditions~\eqref{eq:mm_endpoints_main} are equivalent to\footnote{For an initially stationary oscillator,
the Ermakov variables are related to the quadrature transformation by
\[
M_iM_i^{\mathsf T}
=
\begin{pmatrix}
b_i^2 & b_i\dot b_i/\omega_i \\
b_i\dot b_i/\omega_i &
b_i^{-2}+(\dot b_i/\omega_i)^2
\end{pmatrix}.
\]
Thus, $b_i(t_f)=1$ and $\dot b_i(t_f)=0$ are equivalent to
$M_iM_i^{\mathsf T}=I$. The matrix $M_i$ itself may still be a
phase-space rotation, corresponding to ordinary harmonic phase evolution.}  $M_iM_i^{\mathsf T}=I$: the complete transformation is then a phase-space rotation, so the mode acquires only its ordinary harmonic phase and no squeezing or change of occupation. For a full-cycle palindrome, time-reversal symmetry already enforces $(M_i)_{11}=(M_i)_{22}$. Since $\det M_i=1$, exact refocusing then reduces to the single scalar condition
\begin{equation}
 f_i\equiv(M_i)_{12}+(M_i)_{21}=0
 \label{eq:mm_g_main}
\end{equation}
for each distinct frequency. The proof and the corresponding unrestricted parameter count are given in Appendix~\ref{app:multimode}.


\begin{figure*}[t]
\centering
\begin{tikzpicture}
\begin{groupplot}[
group style={group size=2 by 1,horizontal sep=2.8cm},
width=0.44\textwidth,height=5.2cm,
view={0}{90},
xlabel={$x=\omega_rT$},
xmin=0.2,xmax=1.0,
ymin=2,ymax=6,
colorbar,
colorbar style={font=\scriptsize},
tick label style={font=\small},
label style={font=\small}]

\nextgroupplot[
ylabel={$\eta=\omega_r/\omega_z$},
title={recovery time $\omega_rt_{\rm rec}^{(rz)}$}
]

\addplot[matrix plot*,mesh/cols=41,point meta=explicit]
table[x=x,y=eta,meta=post_gate_total]
{figs/continuation_lambda2.dat};

\addplot[
only marks,
mark=*,
mark size=2.4pt,
draw=black,
fill=white,
line width=0.6pt
]
coordinates {(0.5529,3.75)};

\nextgroupplot[
ylabel={$\eta=\omega_r/\omega_z$},
title={local-uniqueness margin $\sigma_{\min}(\mathcal J)$}
]

\addplot[matrix plot*,mesh/cols=41,point meta=explicit]
table[x=x,y=eta,meta=sigma_min]
{figs/continuation_lambda2.dat};

\addplot[
only marks,
mark=*,
mark size=2.4pt,
draw=black,
fill=white,
line width=0.6pt
]
coordinates {(0.5529,3.75)};

\end{groupplot}
\end{tikzpicture}

\caption{\textbf{Continuation of the exact radial--axial $\Lambda=2$ palindrome.}
The left panel gives the post-gate recovery time of the positive branch.
The right panel gives the smallest singular value of the timing
Jacobian~\eqref{eq:mm_jac_def_main}. A vanishing value would indicate a
fold or loss of local uniqueness.
The branch remains positive and regular over the displayed gate-time
and aspect-ratio range.
The white circle marks the operating point used throughout the paper,
$x=\omega_rT=0.5529$ and
$\eta=\omega_r/\omega_z=3.75$, corresponding to the cesium Rydberg-gate
example of Sec.~\ref{sec:gates}.}
\label{fig:mm_continuation}
\end{figure*}
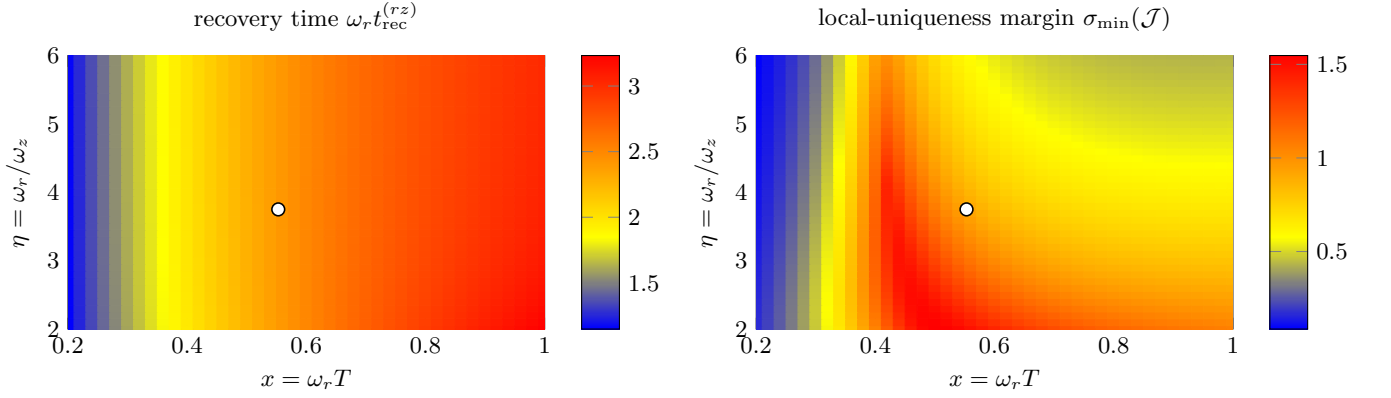

\subsection{Radial--axial closure}

For the two distinct frequencies $\omega_r$ and $\omega_z$, the minimal-parameter palindromic ansatz has two independent durations. Its full chronological sequence is
\begin{equation}
 X(x)\,Y(a)\,X(b)\,Y(a)\,X(x),
 \label{eq:mm_two_main}
\end{equation}
where the first $X(x)$ is the mandatory gate window and the remaining four intervals form the recovery. The complete matrix of mode $i$ is therefore
\begin{equation}
 M_i=F_i(x)R_{i,\Lambda}(a)F_i(b)R_{i,\Lambda}(a)F_i(x),
 \label{eq:mm_two_matrix_main}
\end{equation}
and the two durations are fixed by $f_r=f_z=0$.

At fourfold intensity headroom, $\Lambda=2$, $x=0.5529$, and $\eta=3.75$, the positive solution is
\begin{equation}
a=0.357317...,\qquad
 b=1.12536...
 \label{eq:mm_two_root_main}
\end{equation}
The post-gate recovery time is
\begin{equation}
\omega_r t_{\rm rec}^{(rz)}=2a+b+x
 \approx 2.39,
 \label{eq:mm_two_duration_main}
\end{equation}
which is $4.33\,\mu$s for $\omega_r=2\pi\times88\,$kHz. At nominal depth, an exact alternating four-duration recovery $Y(4.718)X(0.417)Y(2.296)X(0.272)$ gives $\omega_rt_{\rm rec}^{(rz)}=7.70$ ($13.9\,\mu$s for the same frequency; durations rounded for display).

The solution is not an isolated numerical coincidence. Define the residual vector $\bm f(a,b)=(f_r,f_z)^{\mathsf T}$ and its timing Jacobian
\begin{equation}
 \mathcal J\equiv\frac{\partial(f_r,f_z)}{\partial(a,b)}.
 \label{eq:mm_jac_def_main}
\end{equation}
At the root~\eqref{eq:mm_two_root_main}, $\det\mathcal J\approx3.5\neq0$. The implicit-function theorem therefore gives a unique smooth exact-refocusing family under sufficiently small changes of $x$, $\eta$, and $\Lambda$. The smallest singular value, $\sigma_{\min}(\mathcal J)\approx0.92$, measures the distance of the endpoint equations from a singular fold; zero would signal loss of local uniqueness.

Figure~\ref{fig:mm_continuation} continues this $\Lambda=2$ branch over $0.2\le x\le1$ and $2\le\eta\le6$. Positive solutions occur throughout the sampled $41\times41$ grid, with $1.148\le\omega_rt_{\rm rec}^{(rz)}\le3.232$. Moreover, $\sigma_{\min}(\mathcal J)\ge8.6\times10^{-2}$ everywhere on the grid, so every sampled point has a nonsingular local exact family and no fold or loss of positivity is encountered.

\subsection{A fully anisotropic example}

The same idea extends to three distinct frequencies. Consider
\begin{equation}
 (\eta_x,\eta_y,\eta_z)=\left(1,0.95,\frac1{3.75}\right),
 \quad x=0.5529,\quad\Lambda=2,
 \label{eq:mm_aniso_freq_main}
\end{equation}
which represents a $5\%$ radial ellipticity together with the slower axial mode. The three-parameter full palindrome
\begin{equation}
 X(x)Y(a)X(b)Y(c)X(b)Y(a)X(x)
 \label{eq:mm_aniso_sequence_main}
\end{equation}
solves the three endpoint equations $f_x=f_y=f_z=0$ at
\begin{equation}
 a=0.339233...,~~~
 b=1.20259...,~~~
 c=0.429202...
 \label{eq:mm_aniso_root_main}
\end{equation}
Its post-gate recovery time is
\begin{equation}
\omega_rt_{\rm rec}^{(xyz)}=2a+2b+c+x
 \approx4.07,
 \label{eq:mm_aniso_duration_main}
\end{equation}
which is $7.35\,\mu$s at $\omega_r=2\pi\times88\,$kHz. Direct multiplication gives $M_iM_i^{\mathsf T}=I$ for all three axes, 
and the three matrices are rotations. Equivalently, all three Ermakov trajectories satisfy Eq.~\eqref{eq:mm_endpoints_main}; every harmonic occupation distribution is therefore returned exactly, while coherences acquire only known mode-dependent harmonic phases. The corresponding $3\times3$ timing Jacobian is nonsingular ($\det\mathcal J\approx-9.7$), so this fully anisotropic solution also continues as a smooth local family.

\subsection{Multimode optimality bounds}

Each distinct frequency supplies two Ermakov endpoint conditions~\eqref{eq:mm_endpoints_main}. A full-cycle palindrome enforces one of them automatically, leaving one independent scalar equation per frequency. The two- and three-frequency sequences above therefore use the minimum \emph{regular} number of palindromic timing parameters. This is a local parameter-count statement: it does not exclude exceptional singular solutions with fewer controls.

The one-mode minimum-time theorem also gives a rigorous lower bound for any common waveform. Let $J_\Lambda(y)$ denote the certified one-mode minimum recovery time in units of that mode's inverse frequency, as defined in Appendix~\ref{app:pmp}. Then
\begin{equation}
\omega_rt_{\rm rec}\ge
 \max_i\frac{J_\Lambda(\eta_i x)}{\eta_i}.
 \label{eq:mm_lower_bound_main}
\end{equation}
The nominal-depth radial--axial solution lies only $24.9\%$ above this global lower bound. At the $\Lambda=2$ operating point, an exact rational reduction and Sturm-sequence root count further prove that Eq.~\eqref{eq:mm_two_root_main} is the unique shortest nonnegative solution within the two-trap-pulse palindromic family~\eqref{eq:mm_two_main}; the certificate is summarized in Appendix~\ref{app:multimode}. 
Global minimum time over all measurable common controls remains open. Unlike the one-mode problem, different modes can cancel in the multimode Pontryagin switching function, so singular arcs cannot yet be excluded. Multistart searches over longer alternating words found no shorter solution, but this remains numerical evidence rather than a global proof.

\section{Composite echoes for anharmonic traps}
\label{sec:anharm}

The protocol of Sec.~\ref{sec:echo} is exact for harmonic motion, and Sec.~\ref{sec:numerics} showed that in a Gaussian trap the anharmonic remainder scales as $U_0^{-2}$. This is negligible at the deep operating point of Table~\ref{tab:cs}, but dominant in shallower traps. Shallow confinement arises naturally when a fixed laser power is divided among many tweezers or when the register is held in an optical lattice. Here we show that replacing the single hold by a short composite intensity sequence cancels the leading quartic transition amplitudes and changes the residual heating law to $U_0^{-4}$.

The construction can be expressed directly in terms of the Ermakov trajectory. Along the closed piecewise-harmonic cycle, define
\begin{equation}
 \alpha(t)\equiv b(t)e^{i\phi(t)},\qquad \dot\phi(t)=\frac{1}{b^2(t)},
 \label{eq:anh_alpha_main}
\end{equation}
so that the total phase winding is the effective harmonic time $\tau=\int dt/b^2$. Treating the leading anharmonic term $-\beta_4x^4$ perturbatively, the first-order transition amplitude between eigenstates of the \emph{actual} static trap depends on the difference between two moments accumulated during the controlled cycle and the corresponding moments of undisturbed static-trap evolution. In addition to harmonic closure, all first-order quartic population transfer is canceled by
\begin{equation}
\int u(t)\,\alpha^2|\alpha|^2\,dt=e^{i\tau}\sin\tau,
\qquad
\int u(t)\,\alpha^4\,dt=e^{2i\tau}\frac{\sin 2\tau}{2}.
\label{eq:anhcond}
\end{equation}
The integrals extend over the full cycle, and the dark intervals contribute nothing because $u=0$. A derivation from the static-reference error unitary is given in Appendix~\ref{app:quartic}.

The conditions~\eqref{eq:anhcond} are independent of the value of $\beta_4$. The same timing therefore cancels the leading quartic heating of any weakly anharmonic even trap whose leading correction is quartic and which has the same value of $\omega T$; the higher-order coefficients determine the remaining error. The cancellation is operator-level for the off-diagonal first-order generator: it suppresses transitions from every static-trap eigenstate and hence the heating of thermal mixtures without requiring ground-state cooling. It does not, by itself, reproduce the anharmonic phase evolution of arbitrary coherent superpositions. First-order static equivalence of coherences additionally requires the diagonal matching condition $\int u|\alpha|^4dt=\tau$, which is not imposed here.

The full palindromic sequence, now written in physical-time notation to avoid confusion with the dimensionless multimode variables of Sec.~\ref{sec:multimode}, is
\begin{equation}
\begin{aligned}
 \mathrm{off}(T)&\to\mathrm{on}(t_a)\to\mathrm{off}(t_b)\to\mathrm{on}(t_c)\\
 &\to\mathrm{off}(t_b)\to\mathrm{on}(t_a)\to\mathrm{off}(T).
\end{aligned}
 \label{eq:anh_sequence_main}
\end{equation}
The first dark interval is the mandatory gate window and the remaining six intervals form the post-gate recovery. At $\omega T=0.5529$ the solution is
\begin{equation}
(t_a,t_b,t_c)=(0.927799,\;1.030457,\;0.626603)\,\omega^{-1}.
\label{eq:anhroot}
\end{equation}
The harmonic endpoint and the two static-reference moments are satisfied with a maximum numerical residual of $2.1\times10^{-7}$. 
The post-gate recovery time is
\begin{equation}
 t_{\rm rec}^{(4)}=2t_a+2t_b+t_c+T=5.10\,\omega^{-1},
 \label{eq:anh_duration_main}
\end{equation}
about $2.7$ times the single-hold recovery. The maximum Ermakov scale factor is $b_{\max}=1.36$. The sequence uses the same binary intensity hardware as the harmonic echo, although its longer duration and additional switches have their own robustness and latency costs.

\begin{figure}[t]
\centering
\begin{tikzpicture}
\begin{loglogaxis}[width=1.0\linewidth,height=6.1cm,
  xlabel={$U_0\ [\hbar\omega]$},
  ylabel={$\Delta n$ per cycle},
  xmin=25,xmax=480,ymin=2e-9,ymax=3e-3,
  xtick={25, 50, 100, 200, 400},
  xticklabels={25, 50, 100, 200, 400},
  legend style={draw=none,fill=none,font=\scriptsize,at={(0.02,0.02)},anchor=south west},
  legend cell align=left]
\addplot[cGray,mark=square*,thick] table[x=U0,y=harm]{figs/hierarchy_scaling.dat};
\addlegendentry{two-switch echo}
\addplot[cRed,mark=*,very thick] table[x=U0,y=corrected_n0]{figs/corrected_scaling.dat};
\addlegendentry{composite, $n{=}0$}
\addplot[cBlue,mark=triangle*,thick] table[x=U0,y=corrected_n1]{figs/corrected_scaling.dat};
\addlegendentry{composite, $n{=}1$}
\addplot[black,dashed,domain=30:400] {0.4/x^2};
\addplot[black,dotted,thick,domain=30:400] {110/x^4};
\node[font=\scriptsize,rotate=0] at (axis cs:300,9e-6) {$\propto U_0^{-2}$};
\node[font=\scriptsize,rotate=0] at (axis cs:260,7e-8) {$\propto U_0^{-4}$};
\end{loglogaxis}
\end{tikzpicture}
\caption{\textbf{Suppression of anharmonic heating by the composite echo.}
	Heating per cycle versus the depth $U_0$ of the full Gaussian trapping
	potential at $\omega T=0.5529$, obtained by direct wave-function
	propagation. For each value of $U_0$, the initial states are eigenstates
	of the corresponding static Gaussian trap, rather than eigenstates of
	its harmonic approximation. The `two-switch echo' and the
	`composite, $n=0$' calculations start from the Gaussian-trap ground state, while
	the `composite, $n=1$' calculation starts from its first excited eigenstate.
	The composite sequence of
	Eqs.~\eqref{eq:anhcond}--\eqref{eq:anhroot} cancels the complete
	first-order quartic transition amplitude and changes the residual
	heating law from approximately $\Delta n\propto U_0^{-2}$ (dashed line) for the
	two-switch echo to $\Delta n\propto U_0^{-4}$ (dotted) for both tested
	Gaussian-trap eigenstates.}
\label{fig:anharm}
\end{figure}
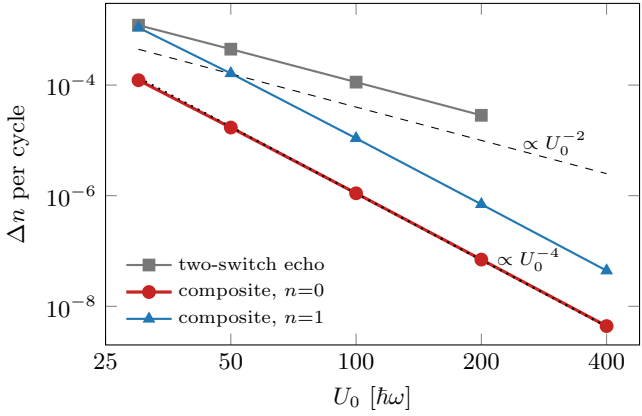

Figure~\ref{fig:anharm} gives the direct wave-packet verification. The fitted depth dependence changes from $U_0^{-1.98}$ for the two-switch echo to $U_0^{-3.96}$ and $U_0^{-3.91}$ for the ground and first excited static-trap eigenstates, respectively, as expected when the complete first-order quartic transition generator is canceled. At $U_0=50\,\hbar\omega$ the composite reduces the heating by a factor of $26$ relative to the two-switch echo; at $U_0=200\,\hbar\omega$ the factor is about $400$. 

\section{Discussion and outlook}
\label{sec:discussion}

Our echo protocol connects several mature bodies of work that had not met in the tweezer-array context. Invariant-based shortcuts to adiabaticity engineer $\omega(t)$ to connect equilibrium states of two \emph{static} traps~\cite{Chen2010,Schaff2010,Schaff2011,GueryOdelin2019}, with bang-bang solutions known to be time optimal~\cite{Stefanatos2010,Stefanatos2011}. Delta-kick cooling and matter-wave lensing remove the position--momentum correlation of an expanding cloud with an impulsive potential~\cite{Chu1986,Ammann1997,Morinaga1999,Kovachy2015,Muntinga2013,Deppner2021}, and the two frameworks were recently unified~\cite{Dupays2021}. In the sudden approximation the release and the catch are frequency jumps, whose squeezing action, and its exact reversal by a suitably delayed second jump, is a classic result for the quantum oscillator~\cite{FanZaidi1988,Kiss1994}, recently demonstrated with trapped atoms~\cite{Xin2021}. Our echo scheme is the finite-duration, intensity-bounded member of this family, timed so that the net squeezing cancels exactly. The two-switch timing at the nominal depth was first obtained, at the level of classical phase-space distributions, in the blinking-tweezer work of Kim, Kim, and Ahn~\cite{KimAhn2025}, where it guarantees stroboscopic survival under time-shared trapping and was confirmed experimentally with single atoms. 

Relative to these works, the present analysis makes several distinct contributions. We establish the global time-optimality theorem (Appendix~\ref{app:pmp}), including its explicit domain of validity for arbitrary intensity headroom. We derive the closed-form single-hold timing~\eqref{eq:t1gen} at arbitrary catch depth and show that the final dark window remains equal to the initial interval, $t_2=T$, throughout this family. The operator identity~\eqref{eq:Ucyc} makes the quantum content of the classical refocusing explicit by showing that each matched harmonic mode undergoes exact unitary evolution under its original trap Hamiltonian. This guarantees the recovery of motional populations and coherences, as verified here by direct wave-function propagation. We further construct common-intensity sequences that refocus several nondegenerate harmonic modes simultaneously, including smooth radial--axial families and a fully anisotropic three-mode example in Sec.~\ref{sec:multimode}. Beyond the harmonic approximation, Sec.~\ref{sec:anharm} derives recovery conditions for weakly anharmonic even traps and gives a palindromic composite echo that cancels the complete first-order quartic transition generator. For a Gaussian tweezer, this changes the asymptotic residual-heating scaling from $\Delta n\propto U_0^{-2}$ for the harmonic timing to $\Delta n\propto U_0^{-4}$. Finally, we apply the refocusing protocol to trap-off Rydberg-gate scheduling and quantify its effect through a mode-resolved heating and Doppler-error budget.

Trap-quench engineering of single-atom motion is already within demonstrated experimental capability. Motional squeezing by tweezer-intensity manipulation has been realized at the $5.8$-dB level~\cite{Lienhard2025}, and controlled tweezer throw-and-catch operations have also been demonstrated~\cite{Hwang2023}. The protocols considered here require only programmable trap-intensity control with nanosecond-scale timing and percent-level intensity accuracy, as discussed in Sec.~\ref{sec:numerics}. In the highest-fidelity processors presently available, trap-off intervals are kept to a few hundred nanoseconds and temperature-related errors form only a small part of the total error budget~\cite{Evered2026}. The present approach is particularly relevant to deeper circuits and to operations requiring longer dark intervals, where repeated release and recapture would otherwise necessitate periodic recooling~\cite{Lin2026}.

Two byproducts are worth noting. First, the protocol furnishes a built-in null test and thermometer: at the analytic point $(t_1,t_2)$ the heating vanishes identically, while detuning $t_1$ produces heating proportional to $(2\bar n+1)$, so scanning the hold time around $t_1$ yields a background-suppressed null measurement of the initial occupation, complementary to release-and-recapture thermometry~\cite{Tuchendler2008,Hoelzl2023,Glatthard2022}. Second, because the cycle acts as pure trap evolution on each atom, it applies unchanged to arrays of tweezers or to optical lattices. Beyond the anharmonic corrections of Sec.~\ref{sec:anharm}, longer composite waveforms offer further capabilities: they can be optimized for robustness against site-to-site frequency spread and can be designed to cancel the residual photon-recoil excitation of Sec.~\ref{sec:gates} together with the breathing excitation. Section~\ref{sec:multimode} shows that slow and fast harmonic modes can already be refocused in parallel with a single global intensity waveform; independent axial control would provide additional speed and robustness.

The combination of the echo with its anharmonic correction also changes which trap architectures are compatible with trap-off gates, with optical lattices providing a particularly attractive example. We denote the dimensionless lattice depth by $s= V_0/E_{\mathrm R}$, with $E_{\mathrm R}$ being the single-photon recoil energy and $V_0$ the lattice depth. Near a lattice minimum, the corresponding harmonic frequency is $\omega=2\sqrt{s}\,E_{\mathrm R}/\hbar$. Optical lattices are highly efficient for holding large atomic registers. A retro-reflected standing wave enhances the trap depth fourfold for a given incident power, thousands of sites can be generated by a single pair of beams, and the interference-defined trapping pattern avoids the site-dependent imaging aberrations that can occur when tweezer arrays are projected through high-numerical-aperture optics~\cite{Brennen1999,Bloch2005}. 

The long-standing drawback of optical lattices as a gate platform is that the entire lattice is formed by one optical mode. Executing a Rydberg gate on any selected subset of atom pairs therefore requires extinguishing the confinement for every atom in the register. With sudden recapture, each such blink heats all spectator atoms by  $\sim7\times10^{-2}$ vibrational quanta at $s=1600$, which becomes prohibitive after a few hundred gate layers. 

Our echo scheme removes this obstacle. Because the control consists of a single global intensity waveform and the cycle is translation invariant, every lattice site is refocused by the same sequence without site-by-site calibration. The composite protocol of Sec.~\ref{sec:anharm} further suppresses the residual anharmonic excitation, which becomes the leading contribution once the harmonic breathing is canceled. For cesium in a $1064$-nm lattice with $s=1600$, corresponding to $V_0/k_{\mathrm B}\simeq102\,\mu\mathrm K$ and $\omega=2\pi\times106\,\mathrm{kHz}$, the value $\omega T=0.55$ corresponds to a gate window of $T=0.83\,\mu\mathrm{s}$. At these parameters, the per-blink excitation decreases from approximately $7\times10^{-2}$ quanta for sudden recapture to $1.3\times10^{-3}$ quanta for the two-switch harmonic echo, and then to $1.3\times10^{-4}$ quanta with the anharmonic composite sequence. Consequently, $10^{3}$--$10^{4}$ gate layers add only about $0.13$--$1.3$ vibrational quantum to each spectator atom. Echo-protected blinking may therefore enable neutral-atom quantum computation with the register held in an optical lattice, combining the lattice's power efficiency and spatial homogeneity with trap-off entangling gates.

\emph{Note added.}---After the completion of this work we became aware of a recent preprint by Magro, Adamczyk, and de L\'es\'eleuc~\cite{Magro2026}, which independently and in parallel develops multi-pulse on/off motion-control protocols for Rydberg gates in a combined tweezer--lattice geometry. The approaches are complementary: Ref.~\cite{Magro2026} targets gate errors from position noise in interaction-driven gates, while the present work targets recapture heating and its anharmonic corrections.

\begin{acknowledgments}

This research was supported by the TWIN-Q Quantum Computing Center of the Israel Council for Higher Education, the Israel Science Foundation (ISF) under Grant No. 3348/25, the Pazy Research Foundation, and partially by the Helen Diller Quantum Center at the Technion.
\end{acknowledgments}

\appendix

\section{Hold time and effective evolution time}
\label{app:tau}

During a hold at $u=\Lambda^2$ the quantity $\varepsilon_\Lambda=\tfrac12\dot b^2+\tfrac12\Lambda^2 b^2+1/(2b^2)$ is conserved, and $s\equiv b^2$ obeys $\dot s^2=-4\Lambda^2s^2+8\varepsilon_\Lambda s-4$, i.e., harmonic motion at the breathing frequency $2\Lambda$:
\begin{equation}
s(t)=\frac{\varepsilon_\Lambda}{\Lambda^2}+\frac{\sqrt{\varepsilon_\Lambda^2-\Lambda^2}}{\Lambda^2}\,
\sin\!\big(2\Lambda t+\varphi_0\big).
\label{eq:soft}
\end{equation}
Entering from the release point~\eqref{eq:release} fixes $\varepsilon_\Lambda=\tfrac12+\tfrac12\Lambda^2(1+T^2)$ and $\sin\varphi_0=\big[\Lambda^2 s(0)-\varepsilon_\Lambda\big]/\sqrt{\varepsilon_\Lambda^2-\Lambda^2}$ with $s(0)=1+T^2$, which is precisely the arcsine argument of Eq.~\eqref{eq:t1gen}; $\dot s(0)>0$ selects $\cos\varphi_0>0$. The hold ends when $s$ first returns to its initial value with $\dot s<0$, i.e., at $2\Lambda t_1+\varphi_0=\pi-\varphi_0$, which is Eq.~\eqref{eq:t1gen}; for $\Lambda=1$, $\sin\varphi_0=T/\sqrt{4+T^2}$, i.e., $\tan\varphi_0=T/2$ and $t_1=\pi/2-\arctan(T/2)$.

For $\Lambda=1$ write $s(t)=E+A\sin(2t+\varphi_0)$ with $E=1+T^2/2$ and $A=\tfrac{T}{2}\sqrt{4+T^2}$, so that $E^2-A^2=1$. The effective evolution time contributed by the hold is
\begin{align}
\tau_1&=\int_0^{t_1}\frac{\dd t}{s(t)}
=\frac12\int_{\varphi_0}^{\pi-\varphi_0}\frac{\dd y}{E+A\sin y}
\nonumber\\[2pt]
&=\arctan\!\Big[E\cot\tfrac{\varphi_0}{2}+A\Big]
-\arctan\!\Big[E\tan\tfrac{\varphi_0}{2}+A\Big],
\label{eq:tau1}
\end{align}
using $\int\dd y/(E+A\sin y)=2\arctan[E\tan(y/2)+A]$ for $E^2-A^2=1$. Each free-flight window contributes $\int_0^T\dd t/(1+t^2)=\arctan T$, giving Eq.~\eqref{eq:Ucyc}. Since $s>1$ throughout the hold arc, $\tau_1<t_1$. The sum admits a remarkably compact closed form: the symplectic identity $F(T)R(t_1)F(T)=R(\pi-t_1)$ at $\tan(\omega t_1)=2/(\omega T)$, with $F$ the free shear and $R$ the trap rotation (the matter-wave $ABCD$ matrices~\cite{Riou2008}), gives directly
\begin{equation}
\omega\tau=\pi-\omega t_1=\frac{\pi}{2}+\arctan\frac{\omega T}{2},
\end{equation}
in agreement with Eq.~\eqref{eq:tau1}. For $\omega T=1$, $t_1=1.107\,\omega^{-1}$ and $\tau=2.034\,\omega^{-1}$, values confirmed by the wave-packet simulations of Sec.~\ref{sec:numerics} to $10^{-8}$.

\section{Time optimality: theorem and limits}
\label{app:pmp}

We minimize the recovery time after the mandatory dark window over all measurable controls $u(t)\in[0,1]$, i.e., the time to steer $(b,\dot b)$ from the release point $R_x=(\sqrt{1+x^2},\,x/\sqrt{1+x^2})$, $x\equiv\omega T$, to $G=(1,0)$ under Eq.~\eqref{eq:ermb}. Write $v=\dot b$, and let $X$ ($u=0$) and $Y$ ($u=1$) denote the free and nominal-trap vector fields. An optimal control exists by Filippov's theorem: the velocity sets are segments (convex and compact), and $W=\tfrac12(v^2+b^2+b^{-2})$ obeys $\dot W=(1-u)\,bv\le W$, so trajectories remain in a compact subset of $\{b>0\}$ on bounded intervals. The related trap-to-trap cooling problem was analyzed with the same machinery in Refs.~\cite{Stefanatos2010,Stefanatos2011} (see Ref.~\cite{Boscain2021} for a tutorial introduction to the maximum principle in quantum control); the theorem below addresses the different boundary conditions of the recovery problem.

\emph{Extremal structure.}---The minimum-time Hamiltonian is $\mathcal H=-1+\lambda_1v+\lambda_2(b^{-3}-ub)$, with adjoints $\dot\lambda_1=(u+3b^{-4})\lambda_2$, $\dot\lambda_2=-\lambda_1$. The coefficient of $u$ is $-b\lambda_2$; since $b>0$ the switching function is $\Phi=-\lambda_2$, with $u=1$ for $\Phi>0$ and $u=0$ for $\Phi<0$, and $\dot\Phi=\lambda_1$ at any zero of $\Phi$. If $\Phi$ and $\dot\Phi$ vanished together, then $\lambda_1=\lambda_2=0$, contradicting $\mathcal H\equiv0$ in the normal case and costate nontriviality in the abnormal case: there are no singular arcs, every optimal control is bang-bang, and all switches are transversal. Abnormal extremals are excluded outright: they require $\lambda_1v=0$ at switches, hence switches only at turning points $v=0$, from which no admissible concatenation reaches $G$ (a free arc started at $v\ge0$ has $\dot v=b^{-3}>0$ and never switches again nor reaches $G$, whose free invariant differs unless the turning point is already $b=1$, impossible for $x>0$). At a normal switch $\mathcal H=0$ gives $\lambda_1v=1$, so switches $X\!\to\!Y$ occur only at $v>0$ and $Y\!\to\!X$ only at $v<0$; and since $G$ is an equilibrium of $Y$, the last arc is always $X$. Zero-switch solutions fail ($X$ from $R_x$ expands monotonically; $Y$ conserves $v^2+b^2+b^{-2}$, which equals $2+x^2$ at $R_x$ but $2$ at $G$), so the admissible structures are $YX$, $XYX$, $YXYX,\dots$

\emph{One and two switches.}---Free arcs conserve $I_0=v^2+b^{-2}$, and both $R_x$ and $G$ lie on $I_0=1$; the contracting branch reaching $G$ is $\Gamma_-=\{(\sqrt{1+s^2},-s/\sqrt{1+s^2})\}$ with $s$ the remaining flight time. Trap arcs conserve $I_1=v^2+b^2+b^{-2}$, equal to $2+x^2$ at $R_x$ and $2+s^2$ on $\Gamma_-$: a $Y$ arc from $R_x$ can meet the final free orbit only at $s=x$. The final free duration is therefore \emph{forced} to equal $x$, an independent derivation of Eq.~\eqref{eq:t2}, and the switching point is the mirror image $R_x^-$, first reached after $\arctan(2/x)=t_1$ [Eq.~\eqref{eq:t1}]. The unique single-hold extremal is the protocol, with recovery time $J_*(x)=x+\arctan(2/x)$. A two-switch extremal must be $XYX$; its initial free arc of duration $a$ merely advances $R_x\to R_{x+a}$, giving $J_{XYX}(a)=x+2a+\arctan[2/(x+a)]$ with $dJ/da=2-2/[(x+a)^2+4]>0$: strictly slower for every $a>0$.

\emph{Interior arcs.}---For $g=(0,-b)$ and $k=(b,-v)$ one has $[h_u,g]=k$ and $[h_u,k]=2h_u-4ug$, where $h_u$ is the controlled field. Along an extremal the pairings $W_g=\lambda\!\cdot\!g$ (proportional to $\Phi$) and $W_k=\lambda\!\cdot\!k$ obey $\dot W_g=W_k$, $\dot W_k=2-4uW_g$, using $\lambda\!\cdot\!h_u=1$. On a free arc that begins at a $Y\!\to\!X$ switch with slope $-r$ ($r=|v|/b$), $W_g=t^2-t/r$: the next switch occurs at $t_X=1/r$, and the flow maps $(b,-rb)\mapsto(1/rb,\,1/b)$, reversing the slope to $+r$. On a trap arc that begins at an $X\!\to\!Y$ switch with slope $+r$, $W_g=\sin t\,(\sin t+\cos t/r)$: the next switch occurs at $t_Y=\pi/2+\arctan r$, with exit slope $-r$. Every interior pair of consecutive arcs therefore lasts $1/r+\pi/2+\arctan r>\pi$, since $\arctan(1/r)<1/r$.

\emph{Theorem.}---$J_*(x)$ is strictly increasing and $J_*(x)\le\pi$ for $x\le x_{\rm clean}=2.4587\ldots$, so no extremal with three or more switches can compete there. The bound sharpens by including the final approach: the last three arcs of any such extremal are $X,Y,X$ with the final free duration $d$ fixing $r=d/(1+d^2)\le\tfrac12$, so its total time is at least $L(d)=2d+1/d+\pi/2+\arctan[d/(1+d^2)]$, minimized at $2d_0^6+4d_0^4=1$, $d_0=0.6720\ldots$, giving $L_{\min}=4.8365\,\omega^{-1}$. Hence, for
\begin{equation}
0<\omega T\le x_{\rm cert}=4.4107\ldots,
\qquad J_*(x_{\rm cert})=L_{\min},
\label{eq:xcert}
\end{equation}
the two-switch protocol of Eqs.~\eqref{eq:t1} and \eqref{eq:t2} is the unique globally time-optimal recovery, up to zero-duration arcs and control values at isolated switching instants. Because every interior pair lasts more than $\pi$, extremals of bounded duration contain finitely many switches, so chattering is excluded. Direct numerical integration of the state and costate equations reproduces the switching relations and endpoint conditions above with residuals below $10^{-9}$.

\emph{Failure at large $T$.}---No unrestricted claim holds. As an example, at $x=10$ the four-switch extremal with durations $(0.408784,\,3.121215,\,1.880851,\,2.758729)\,\omega^{-1}$ for $Y,X,Y,X$ reaches $G$ exactly, 
recovering in $8.170\,\omega^{-1}$ instead of $J_*(10)=10.197\,\omega^{-1}$ 
by transiently compressing the packet to $b\approx0.32$ and thereby avoiding the long final coast. Numerical continuation shows this family first ties the two-switch protocol at $\omega T\approx6.902$; between $x_{\rm cert}$ and this point optimality is uncertified but no faster solution was found numerically.

\emph{Intensity headroom.}---The analysis extends to $0\le u(t)\le\Lambda^2$ with one structural change: $G$ is no longer an equilibrium of the strong-trap field $Y$ ($u=\Lambda^2$), so extremals may terminate with a $Y$ arc. The maximum-principle structure (no singular arcs, bang-bang controls, and the switch-direction rules) carries over verbatim. Abnormal extremals are again excluded, now by a short argument: every abnormal switch must occur at $v=0$; from the release point an initial $X$ arc has $v>0$ throughout and never switches, while an initial $Y$ arc can switch only at its turning point, after which the free acceleration is positive and no further abnormal switch is possible, and the free-flight invariant excludes arrival at $G$; a switch-free $Y$ arc is excluded because the strong-trap invariant differs between the release point and $G$. The unique single-hold extremal is the single-hold echo $Y(t_1)X(T)$ with $t_1$ given by Eq.~\eqref{eq:t1gen}, equivalently $t_1=\Lambda^{-1}\arctan\{2\Lambda x/[\Lambda^2(1+x^2)-1]\}$; its recovery time $J_\Lambda(x)=x+t_1$ is strictly increasing in $x$, and every $XYX$ extremal is again strictly slower. Interior arcs last $t_X=1/r$ and $t_Y=\Lambda^{-1}[\pi/2+\arctan(r/\Lambda)]$. Extremals with three or more switches ending in free flight are bounded below by $C_X(\Lambda)$, the minimum over $0<d\le1$ of $2d+1/d+\Lambda^{-1}[\pi/2+\arctan(d/\{\Lambda(1+d^2)\})]$, attained at the positive root of $2\Lambda^2d^6+(3\Lambda^2+1)d^4=\Lambda^2$; extremals ending in a strong-trap arc are bounded below by
\begin{equation}
C_Y(\Lambda)=\frac{1}{(\Lambda-1)\sqrt{\Lambda}}+\frac{\arctan\sqrt{\Lambda}}{\Lambda},
\end{equation}
obtained by minimizing the final free-plus-trap block over the backward orbit from $G$: the switch slope on the backward $Y$ orbit is $r(z)=\Lambda(\Lambda^2-1)z/(\Lambda^2+z^2)$ with $z=\tan(\Lambda t_Y)$, and minimizing $1/r(z)+\arctan(z)/\Lambda$ gives $z=\sqrt{\Lambda}$, from which the display above follows. The single-hold echo is therefore the unique global time optimum whenever $J_\Lambda(x)<C(\Lambda)\equiv\min\{C_X(\Lambda),C_Y(\Lambda)\}$; the certified boundary $J_\Lambda(x_{\rm cert})=C(\Lambda)$ evaluates to $x_{\rm cert}=4.4107$, $1.854$, $0.9234$, and $0.5392$ for $\Lambda=1$, $1.5$, $2$, and $3$ ($C_Y=\infty$ at $\Lambda=1$, recovering the theorem above). 
The terminal-$Y$ branch is the one that eventually wins: at the operating point $x=0.5529$ of Sec.~\ref{sec:gates}, overcompression sequences overtake the direct echo near $\Lambda\approx5.61$ (numerically). At $\Lambda=6$, for example, $Y(0.18765)X(0.10102)Y(0.25320)$ reaches $G$ in $0.5419\,\omega^{-1}$, beating even the dark-window floor $T$.

\section{Multimode harmonic echoes}
\label{app:multimode}

This appendix gives the proofs and numerical certificates underlying Sec.~\ref{sec:multimode}. All segment variables $x,a,b,c,\ldots$ are dimensionless durations in units of $\omega_r^{-1}$. We use the frequency ratios $\eta_i=\omega_i/\omega_r$, the normalized quadrature vector $\hat{\bm z}_i$, and the matrices $F_i$ and $R_{i,\Lambda}$ defined in Eqs.~\eqref{eq:mm_eta_main}--\eqref{eq:mm_elements_main}. Matrix products act on column quadratures, with the earliest propagator on the right.

\subsection{Palindromic reduction and local existence}
Let $\Sigma=\operatorname{diag}(1,-1)$, which implements time reversal in the normalized quadrature space,
$\Sigma(\hat Q_i,\hat P_i)^{\mathsf T}
=(\hat Q_i,-\hat P_i)^{\mathsf T}$.
The free and trap-on propagators defined in
Eqs.~\eqref{eq:mm_eta_main}--\eqref{eq:mm_elements_main}
satisfy
\begin{equation}
	\Sigma F_i(s)\Sigma=F_i^{-1}(s),
	\qquad
	\Sigma R_{i,\Lambda}(s)\Sigma
	=R_{i,\Lambda}^{-1}(s).
	\label{eq:mm_element_reversal_app}
\end{equation}
For a full-cycle palindromic intensity waveform, the ordered product
$M_i$ therefore obeys
\begin{equation}
	M_i=\Sigma M_i^{-1}\Sigma.
	\label{eq:mm_reversal_app}
\end{equation}
Writing
\[
M_i=
\begin{pmatrix}
	A_i&B_i\\
	C_i&D_i
\end{pmatrix},
\]
Eq.~\eqref{eq:mm_reversal_app} implies $A_i=D_i$. Since every
segment matrix is symplectic, $\det M_i=1$. Exact closure of the
$i$th mode is consequently equivalent to the single scalar condition
\begin{equation}
	f_i\equiv B_i+C_i
	=(M_i)_{12}+(M_i)_{21}=0.
	\label{eq:mm_scalar_closure_app}
\end{equation}
Indeed, $A_i=D_i$, $B_i=-C_i$, and $\det M_i=1$ imply
$M_iM_i^{\mathsf T}=I$, which is equivalent to the Ermakov endpoint
conditions
\[
b_i(t_f)=1,\qquad \dot b_i(t_f)=0.
\]

Without the palindromic symmetry, two independent residuals are
required for each mode:
\begin{equation}
	r_{i,1}=(M_i)_{11}-(M_i)_{22},
	\qquad
	r_{i,2}=(M_i)_{12}+(M_i)_{21}.
\end{equation}
A regular endpoint map for $q$ distinct frequencies therefore requires
at least $q$ independent durations within the palindromic class, or
$2q$ durations for an unrestricted sequence. This is a local rank
statement and does not exclude exceptional singular solutions with
fewer controls.

Let
\[
\bm s=(s_1,\ldots,s_q)
\]
denote the independent palindromic durations, and let $\bm p$ collect
the fixed parameters $x=\omega_rT$, the frequency ratios $\eta_i$, and
the available catch depth $\Lambda$. We define the vector of closure
residuals by
\[
\bm f(\bm s;\bm p)
=
\bigl(f_1,\ldots,f_q\bigr)^{\mathsf T}.
\]
If an exact solution $(\bm s_0,\bm p_0)$ satisfies
\begin{equation}
	\bm f(\bm s_0;\bm p_0)=0,
	\qquad
	\det\mathcal J\neq0,
	\qquad
	\mathcal J_{ij}
	=
	\left.
	\frac{\partial f_i}{\partial s_j}
	\right|_{(\bm s_0,\bm p_0)},
	\label{eq:mm_ift_app}
\end{equation}
then the implicit-function theorem guarantees a unique smooth family
of exact solutions $\bm s=\bm s(\bm p)$ in a neighborhood of
$\bm p_0$. Its parameter dependence is
\begin{equation}
	\frac{\partial\bm s}{\partial p_\mu}
	=
	-\mathcal J^{-1}
	\frac{\partial\bm f}{\partial p_\mu}.
\end{equation}

For completeness, consider a general sequence containing $N$
post-release segments. Denote the propagator of segment $j$ for mode
$i$ by
\begin{equation}
	E_{i,j}(s_j)=
	\begin{cases}
		F_i(s_j),&\text{if segment $j$ is trap off},\\[2pt]
		R_{i,\Lambda}(s_j),&\text{if segment $j$ is trap on}.
	\end{cases}
	\label{eq:mm_segment_matrix_app}
\end{equation}
The complete cycle matrix can then be written as
\begin{equation}
	M_i=
	E_{i,N}(s_N)\cdots E_{i,2}(s_2)E_{i,1}(s_1)F_i(x),
	\label{eq:mm_general_product_app}
\end{equation}
where the earliest propagator appears on the right. The derivative
with respect to a segment duration is
\begin{equation}
	\frac{\partial M_i}{\partial s_j}
	=
	E_{i,N}\cdots E_{i,j+1}
	A_{i,j}E_{i,j}
	E_{i,j-1}\cdots E_{i,1}F_i(x),
	\label{eq:mm_product_derivative_app}
\end{equation}
with the constant segment generator
\begin{equation}
	A_{i,j}=
	\begin{cases}
		A_{X,i}
		=\eta_i
		\begin{pmatrix}
			0&1\\
			0&0
		\end{pmatrix},
		&E_{i,j}=F_i,\\[12pt]
		A_{Y,i}
		=\eta_i
		\begin{pmatrix}
			0&1\\
			-\Lambda^2&0
		\end{pmatrix},
		&E_{i,j}=R_{i,\Lambda}.
	\end{cases}
	\label{eq:mm_generators_app}
\end{equation}
These expressions provide the analytic derivatives used to evaluate
the timing Jacobian $\mathcal J$.

For the radial--axial root~\eqref{eq:mm_two_root_main},
\begin{equation}
 \mathcal J=\begin{pmatrix}
 3.13690164445&1.71882482946\\
 -1.54907123514&0.266951544665
 \end{pmatrix},
\end{equation}
which gives the determinant and smallest singular value quoted in the main text. For the fully anisotropic root~\eqref{eq:mm_aniso_root_main},
\begin{equation}
 \mathcal J=\begin{pmatrix}
 8.15711523&-0.92491169&-5.22730595\\
 8.38297472&-0.41353449&-4.39785858\\
 -1.13540580&0.47258912&-0.86475617
 \end{pmatrix},
\end{equation}
with $\det\mathcal J=-9.70734257$ and $\sigma_{\min}(\mathcal J)=0.48721249$. If the repeated palindrome durations are released and treated independently, the full endpoint Jacobians remain nonsingular: $\det J_{4\times4}=2.89227257$, $\sigma_{\min}=0.17209107$ for the radial--axial word, and $\det J_{6\times6}=-5.23442418$, $\sigma_{\min}=0.20782411$ for the fully anisotropic word. The displayed solutions therefore attain the regular parameter-count bounds.

\subsection{Analytic reduction and continuation}

For the two-mode palindrome~\eqref{eq:mm_two_main}, set $z_i=\tan(\Lambda\eta_i a)$. The scalar closure equation is linear in the intermediate dark duration $b$. Defining
\begin{align}
 A_i(a)&=\big[\Lambda^2(1+\eta_i^2x^2)-1\big]z_i
 +\Lambda\eta_i x(z_i^2-1),\\
 D_i(a)&=\Lambda^2(1+\eta_i^2x^2)z_i^2
 -2\Lambda\eta_i xz_i+1,
\end{align}
we obtain
\begin{equation}
 b=B_i(a)\equiv\frac{2A_i(a)}{\Lambda\eta_iD_i(a)}.
 \label{eq:mm_B_app}
\end{equation}
For the cylindrically symmetric radial--axial trap, the common-control problem is the one-dimensional equation $B_r(a)=B_z(a)$, equivalently $B_{\eta_i=1}(a)=B_{\eta_i=1/\eta}(a)$, followed by $b=B_r(a)$. This reduction was used for Fig.~\ref{fig:mm_continuation}. The continuation uses bounded root polishing at every point, analytic Jacobians, and endpoint residuals below $1.7\times10^{-15}$. Continuing the fully anisotropic branch as $\eta_y$ varies from $0.90$ to $0.99$ changes $\omega_rt_{\rm rec}^{(xyz)}$ smoothly from $4.12638$ to $4.01553$ without loss of regularity.

\subsection{Bounds and exact optimality within the palindromic family}

Let
\begin{align}
 h_\Lambda(y)&=\frac1\Lambda\operatorname{atan2}\!\left(2\Lambda y,\Lambda^2(1+y^2)-1\right),\\
 J_\Lambda(y)&=y+h_\Lambda(y)
\end{align}
be the certified one-mode minimum recovery time in that mode's natural units. A common waveform of physical duration $t_{\rm rec}$ is admissible for every individual mode. Therefore
\begin{equation}
 \omega_rt_{\rm rec}\ge\max_i\frac{J_\Lambda(\eta_i x)}{\eta_i},
\end{equation}
provided every $\eta_i x$ lies in the certified region of Appendix~\ref{app:pmp}. For the nominal-depth radial--axial root, $\omega_rt_{\rm rec}=7.70364$ and the lower bound is $6.16744$; at $\Lambda=2$, the corresponding values are $2.39290$ and $0.906852$.

For the $\Lambda=2$ two-pulse palindrome at the operating point, take the exact rational parameters
\begin{equation}
 x=\frac{5529}{10000},\qquad \eta=\frac{15}{4},\qquad \eta_z=\frac{4}{15},
 \label{eq:mm_rational_parameters_app}
\end{equation}
and introduce
\begin{equation}
 t=\tan\!\left(\frac{2a}{15}\right),
 \qquad
 a(t)=\frac{15}{2}\arctan t .
 \label{eq:mm_t_definition_app}
\end{equation}
To define the resulting algebraic equation without listing long polynomial coefficients, let $\mathcal N_n(t)$ and $\mathcal D_n(t)$ be the integer polynomials generated by
\begin{align} \label{eq:mm_tangent_recurrence_app}
 \mathcal N_0(t)&=0,& \mathcal D_0(t)&=1,\\
 \mathcal N_{n+1}(t)&=\mathcal N_n(t)+t\mathcal D_n(t),&
 \mathcal D_{n+1}(t)&=\mathcal D_n(t)-t\mathcal N_n(t).\nonumber
\end{align}
The tangent addition formula then gives
\begin{equation}
 \tan\!\bigl(n\arctan t\bigr)
 =\frac{\mathcal N_n(t)}{\mathcal D_n(t)}
 \label{eq:mm_tangent_ratio_app}
\end{equation}
whenever $\mathcal D_n(t)\neq0$. In particular,
\begin{equation}
 \tan(2a)=\frac{\mathcal N_{15}(t)}{\mathcal D_{15}(t)},
 \qquad
 \tan\!\left(\frac{8a}{15}\right)
 =\frac{\mathcal N_4(t)}{\mathcal D_4(t)}.
 \label{eq:mm_tangent_substitutions_app}
\end{equation}
Substituting Eq.~\eqref{eq:mm_tangent_substitutions_app} into the two functions $B_{\eta_i=1}(a)$ and $B_{\eta_i=4/15}(a)$ defined by Eq.~\eqref{eq:mm_B_app} makes their difference a rational function of $t$. After combining the fractions and dividing the numerator and denominator by their polynomial greatest common divisor, define the coprime primitive integer polynomials $P_{37}$ and $Q_{38}$ by
\begin{equation}
 B_{\eta_i=1}\bigl(a(t)\bigr)
 -B_{\eta_i=4/15}\bigl(a(t)\bigr)
 =\frac{tP_{37}(t)}{Q_{38}(t)}.
 \label{eq:mm_rational_closure_app}
\end{equation}
Here $\deg P_{37}=37$ and $\deg Q_{38}=38$. Their remaining overall signs are fixed by requiring the leading coefficient of $Q_{38}$ to be positive. The positive zeros of $P_{37}$ for which $Q_{38}\neq0$ are precisely the nontrivial candidate closure points in this parametrization. Zeros of $Q_{38}$ are poles of the rational closure equation and are not solutions.

For a real polynomial $R$, let $V_R(c)$ denote the number of sign variations in its Sturm chain evaluated at $t=c$, with zeros omitted. Sturm's theorem gives
\begin{equation}
 V_{P_{37}}(0)-V_{P_{37}}(31/250)=2,
 \quad
 V_{Q_{38}}(0)-V_{Q_{38}}(31/250)=0.
 \label{eq:mm_sturm_counts_app}
\end{equation}
Thus $P_{37}$ has exactly two positive roots in $0<t<31/250$, and the rational closure function has no pole in this interval. The two roots give
\begin{align}
 (a,b)&=(0.0504781823324,-0.803656486748),\\
 (a,b)&=(0.357316692864,\phantom{-}1.125364460527).
\end{align}
The first root is inadmissible because $b<0$. Any nonnegative palindrome faster than Eq.~\eqref{eq:mm_two_duration_main} would require $a<0.919998923128$, corresponding through Eq.~\eqref{eq:mm_t_definition_app} to $t<0.123285507<31/250$. Equations~\eqref{eq:mm_rational_closure_app} and \eqref{eq:mm_sturm_counts_app} therefore prove that Eq.~\eqref{eq:mm_two_root_main} is the unique globally shortest solution within the symmetric two-trap-pulse family.

A global multimode minimum-time theorem over all common controls remains open. In Ermakov variables,
\begin{equation}
 \dot b_i=v_i,\qquad \dot v_i=\eta_i^2(b_i^{-3}-ub_i),
\end{equation}
and the Pontryagin switching function is
\begin{equation}
 \Phi=-\sum_i\eta_i^2q_i b_i.
\end{equation}
Unlike the one-mode case, different modal terms can cancel in $\Phi$ and $\dot\Phi$. When its denominator is nonzero, a candidate singular control is
\begin{equation}
 u_s=-\frac{\sum_i(\eta_i^4q_i b_i^{-3}+\eta_i^2p_i v_i)}{\sum_i\eta_i^4q_i b_i},
\end{equation}
which may lie inside the admissible interval. Thus singular arcs are not excluded. Constrained multistart searches over alternating words with five through eight post-gate arcs, beginning with either control value, found no solution shorter than Eq.~\eqref{eq:mm_two_duration_main}; the best solutions collapsed surplus arcs and reproduced the palindrome. This is supporting numerical evidence, not a global proof.

\section{First-order quartic composite conditions}
\label{app:quartic}

We derive the static-reference conditions used in Sec.~\ref{sec:anharm}. In harmonic units let
\begin{equation}
 H_{\rm s}=H_{\rm h}-\beta_4x^4+\cdots,\qquad
 H_{\rm h}=\frac{p^2+x^2}{2},
\end{equation}
where $H_{\rm s}$ is the static weakly anharmonic trap. Let $M(t)$ be the symplectic propagator generated by the piecewise-quadratic control, and define
\begin{equation}
 \alpha(t)=M_{11}(t)+iM_{12}(t)=b(t)e^{i\phi(t)}.
\end{equation}
Equation~\eqref{eq:ermakov} and the unit determinant of $M$ give $\dot\phi=1/b^2$. In the interaction picture of the controlled harmonic evolution,
\begin{equation}
 \widetilde x(t)=\frac{\overline{\alpha(t)}a+\alpha(t)a^\dagger}{\sqrt2}.
 \label{eq:quartic_x_app}
\end{equation}
If the harmonic part of the complete cycle is the rotation $e^{-iH_{\rm h}\tau}$, the actual cycle and the appropriate static reference are, to first order in $\beta_4$,
\begin{align}
 U_{\rm cyc}&=e^{-iH_{\rm h}\tau}\left[I+i\beta_4G_{\rm cyc}+O(\beta_4^2)\right],\\
 e^{-iH_{\rm s}\tau}&=e^{-iH_{\rm h}\tau}\left[I+i\beta_4G_{\rm stat}+O(\beta_4^2)\right],
\end{align}
with
\begin{align}
 G_{\rm cyc}&=\int u(t)\,\widetilde x^4(t)\,dt,\\
 G_{\rm stat}&=\int_0^\tau
 \left[e^{iH_{\rm h}s}xe^{-iH_{\rm h}s}\right]^4ds.
\end{align}
The first-order transition amplitude between distinct eigenstates of $H_{\rm s}$ is therefore governed by the off-diagonal part of $G_{\rm cyc}-G_{\rm stat}$, rather than by $G_{\rm cyc}$ alone.

Normal ordering Eq.~\eqref{eq:quartic_x_app} shows that all terms
changing the harmonic occupation by two quanta carry the common
coefficient functional $\alpha^2|\alpha|^2$, whereas all terms
changing it by four quanta carry $\alpha^4$. We therefore define the
two quartic moments accumulated during the controlled cycle as
\begin{equation}
	C^{\rm cyc}_{4,j}
	\equiv
	\int_0^{t_f}
	u(t)\,
	\alpha^{2j}(t)\,
	|\alpha(t)|^{2(2-j)}
	\,\dd t,
	\qquad j=1,2.
	\label{eq:quartic_cycle_moments_app}
\end{equation}
Explicitly,
\begin{equation}
	C^{\rm cyc}_{4,1}
	=
	\int_0^{t_f}u(t)\alpha^2(t)|\alpha(t)|^2\,\dd t,
	\,\,
	C^{\rm cyc}_{4,2}
	=
	\int_0^{t_f}u(t)\alpha^4(t)\,\dd t.
	\label{eq:quartic_cycle_moments_explicit_app}
\end{equation}

For uninterrupted evolution in the static harmonic reference,
$\alpha_{\rm stat}(s)=e^{is}$. The corresponding reference moments
are defined by
\begin{equation}
	C^{\rm stat}_{4,j}(\tau)
	\equiv
	\int_0^\tau e^{2ijs}\,\dd s,
	\qquad j=1,2.
	\label{eq:quartic_static_moments_app}
\end{equation}
Their explicit values are
\begin{align}
	C^{\rm stat}_{4,1}(\tau)
	&=
	\int_0^\tau e^{2is}\,\dd s
	=
	e^{i\tau}\sin\tau,
	\label{eq:quartic_C41_app}\\
	C^{\rm stat}_{4,2}(\tau)
	&=
	\int_0^\tau e^{4is}\,\dd s
	=
	e^{2i\tau}\frac{\sin 2\tau}{2}.
	\label{eq:quartic_C42_app}
\end{align}

The complete first-order quartic transition generator is canceled
when the cycle moments match the static-reference moments,
\begin{equation}
	C^{\rm cyc}_{4,j}
	=
	C^{\rm stat}_{4,j}(\tau),
	\qquad j=1,2,
	\label{eq:quartic_matching_app}
\end{equation}
which is precisely Eq.~\eqref{eq:anhcond} of the main text. The diagonal normal-ordered sector is proportional to a polynomial in $a^\dagger a$ and is matched to static evolution only if
\begin{equation}
 \int u(t)|\alpha(t)|^4dt=\tau.
 \label{eq:quartic_diagonal_app}
\end{equation}
Equation~\eqref{eq:quartic_diagonal_app} is unnecessary for suppressing population transfer from a static-trap eigenstate, but it is required for first-order equivalence of arbitrary motional coherences.

For a full-cycle palindrome, time reversal gives
\begin{equation}
 \alpha(t_f-t)=e^{i\tau}\overline{\alpha(t)}.
\end{equation}
The cycle and static moments therefore obey the same phase relation, so each complex equality in Eq.~\eqref{eq:anhcond} reduces to one real condition. Together with harmonic closure, three independent durations are sufficient for the quartic no-heating problem. At $\omega T=0.5529$, bounded multistart root finding gives the full-precision palindromic durations
\begin{equation}
 \begin{aligned}
 t_a&=0.927798714802\,\omega^{-1},\\
 t_b&=1.030457248899\,\omega^{-1},\\
 t_c&=0.626603200758\,\omega^{-1}.
 \end{aligned}
 \label{eq:quartic_root_app}
\end{equation}
The maximum residual of the harmonic closure and static-reference moment equations is $2.1\times10^{-7}$ at the highest quadrature order used. Halving the quadrature spacing changes each reported moment by less than $3\times10^{-8}$.

\section{Quantum release and recapture}
\label{app:rr}

Although our motivation is the gate cycle, the scaling solution of Sec.~\ref{sec:rr} also settles a diagnostic question. The survival of an atom after a variable dark time is the standard R\&R thermometer for single atoms~\cite{Tuchendler2008}. To describe it, truncate the trap at its finite depth $U$ and approximate the bound eigenstates $\phi_j$ by those of the ideal oscillator with $E_j<U$; if the trap is restored after a time $t$ and the survival is measured long afterwards, only the projection onto bound states is recaptured,
\begin{equation}
P_{1\mathrm d}(t)=\sum_{\{j\,|\,E_j<U\}}\big|\langle \phi_j|\psi_n(t)\rangle\big|^2 ,
\label{eq:P1d}
\end{equation}
with $\psi_n(t)$ the freely expanded eigenstate of Eq.~\eqref{eq:psin}. In three dimensions the free propagator factorizes and, measuring time in units of $\omega_z^{-1}$ with the auxiliary ratios $\eta_i=\omega_i/\omega_z$,
\begin{equation}
\Psi_{n_x n_y n_z}(t,\bar r)=\psi_{n_x}(\eta_x t,x)\,\psi_{n_y}(\eta_y t,y)\,\psi_{n_z}(t,z),
\label{eq:3dfact}
\end{equation}
so that $P_{3\mathrm d}(t)$ is a sum of products of one-dimensional overlaps, restricted by the three-dimensional bound-state condition $E_{ijk}<U$.

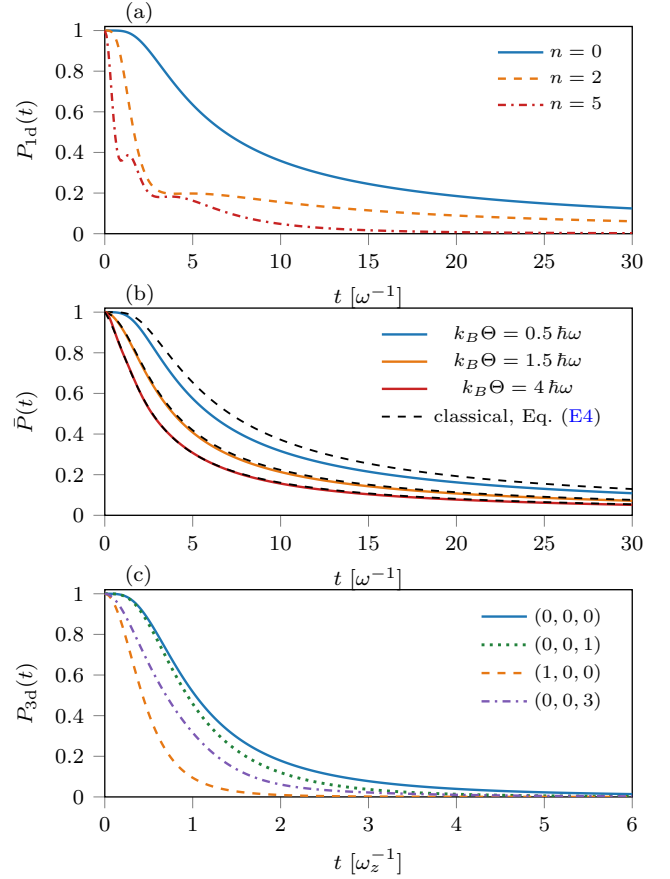
\begin{figure}[tb]
\centering
\begin{tikzpicture}
\begin{groupplot}[group style={group size=1 by 3, vertical sep=28pt},
 rrfig, height=0.5\columnwidth, clip mode=individual,]
\nextgroupplot[xlabel={$t\ [\omega^{-1}]$}, ylabel={$P_{1\mathrm d}(t)$},
  xmin=0,xmax=30,ymin=0,ymax=1.02,
  legend pos=north east]
\addplot[cBlue] table[x=t,y=P0]{figs/p1d.dat};
\addlegendentry{$n=0$}
\addplot[cOrange,dashed] table[x=t,y=P2]{figs/p1d.dat};
\addlegendentry{$n=2$}
\addplot[cRed,dashdotted] table[x=t,y=P5]{figs/p1d.dat};
\addlegendentry{$n=5$}
\node[anchor=south west,font=\footnotesize] at (rel axis cs:0.02,0.97) {(a)};
\nextgroupplot[xlabel={$t\ [\omega^{-1}]$}, ylabel={$\bar P(t)$},
  xmin=0,xmax=30,ymin=0,ymax=1.02,
  legend pos=north east]
\addplot[cBlue] table[x=t,y=Q05]{figs/thermal.dat};
\addlegendentry{$k_B\Theta=0.5\,\hbar\omega$}
\addplot[black, dashed, line width=0.7pt, forget plot] table[x=t,y=C05]{figs/thermal.dat};
\addplot[cOrange] table[x=t,y=Q15]{figs/thermal.dat};
\addlegendentry{$k_B\Theta=1.5\,\hbar\omega$}
\addplot[black, dashed, line width=0.7pt, forget plot] table[x=t,y=C15]{figs/thermal.dat};
\addplot[cRed] table[x=t,y=Q40]{figs/thermal.dat};
\addlegendentry{$k_B\Theta=4\,\hbar\omega$}
\addplot[black, dashed, line width=0.7pt] table[x=t,y=C40]{figs/thermal.dat};
\addlegendentry{classical, Eq.~\eqref{eq:Pclth}}
\node[anchor=south west,font=\footnotesize] at (rel axis cs:0.02,0.97) {(b)};
\nextgroupplot[xlabel={$t\ [\omega_z^{-1}]$}, ylabel={$P_{3\mathrm d}(t)$},
  xmin=0,xmax=6,ymin=0,ymax=1.02,
  legend pos=north east]
\addplot[cBlue] table[x=t,y=P000]{figs/p3d.dat};
\addlegendentry{$(0,0,0)$}
\addplot[cGreen,dotted, line width=1.1pt] table[x=t,y=P001]{figs/p3d.dat};
\addlegendentry{$(0,0,1)$}
\addplot[cOrange,dashed] table[x=t,y=P100]{figs/p3d.dat};
\addlegendentry{$(1,0,0)$}
\addplot[cPurple,dashdotted] table[x=t,y=P003]{figs/p3d.dat};
\addlegendentry{$(0,0,3)$}
\node[anchor=south west,font=\footnotesize] at (rel axis cs:0.02,0.97) {(c)};
\end{groupplot}
\end{tikzpicture}
\caption{\textbf{Quantum release and recapture versus the classical benchmark.}
(a) One-dimensional recapture probability, Eq.~\eqref{eq:P1d}, for a truncated harmonic trap supporting $N=6$ bound states, with the initial state the eigenstate $\phi_n$, $n=0,2,5$. Although the ``temperature'' is zero in all three cases, the signal decays; the $n=5$ signal is non-monotonic, with revivals arising from the multi-lobed structure of the expanding wave function.
(b) Thermal ensembles: Boltzmann-weighted quantum recapture (solid) versus the classical prediction, Eq.~\eqref{eq:Pclth}, at the same temperature (dashed), for $k_B\Theta/\hbar\omega=0.5$, $1.5$, $4$ (mean occupations of the truncated six-level ensembles shown, $\bar n\approx0.16$, $0.94$, $1.8$). The two descriptions converge at high temperature and separate below $k_B\Theta\sim\hbar\omega$, where the classical curve overestimates the survival.
(c) Three-dimensional recapture, Eq.~\eqref{eq:3dfact}, for a cylindrically symmetric trap with $\eta_x=\eta_y=3$ and depth $U=12\hbar\omega_z$, for the initial eigenstates $(n_x,n_y,n_z)$ indicated. The decay is dominated by the radial expansion, and the equal-energy pair $(1,0,0)$ and $(0,0,3)$ decay similarly despite their different spatial structure.}
\label{fig:rr}
\end{figure}

Figure~\ref{fig:rr}(a) shows the resulting recapture curves. Even at zero temperature the signal decays, with a rate set by the number of bound states, and excited eigenstates display strikingly non-classical behavior: multiple decay time scales for $n=2$ and clear revivals for $n=5$, caused by transient overlaps between side lobes of the expanding wave function and those of the bound states. Leakage out of the truncated bound set (the model's continuum) turns on immediately but at high polynomial order, $1-P_{1\mathrm d}\propto t^{2\lceil(N-n)/2\rceil}$ at short times ($\propto t^{6}$ for the ground state here), which is why the low-$n$ signals are initially so flat.

The standard analysis of R\&R thermometry~\cite{Tuchendler2008} is classical, using a Monte Carlo average over a Boltzmann phase-space distribution of point particles. In the present model its prediction takes a closed form. Free flight shears phase space, $x\to x+vt$ at constant $v$, and a particle is recaptured iff it remains inside the escape disk $E'=\tfrac12 v^2+\tfrac12(x+vt)^2<U$. Decomposing the truncated Boltzmann distribution into shells of fixed energy $E$, uniform in the oscillation phase $\psi$ (the one-dimensional density of states is constant), the recaptured energy on a shell is $E'=E\big[1+t^2/2+\tfrac{t}{2}\sqrt{4+t^2}\,\sin\psi\big]$, whose phase average $E(1+t^2/2)$ is the sudden-recapture heating law of Eq.~\eqref{eq:naive}, so the recaptured fraction of each shell is
\begin{equation}
P_{\rm cl}(t;E)=\frac12+\frac1\pi\arcsin\!\left[\frac{U/E-1-t^2/2}{\tfrac{t}{2}\sqrt{4+t^2}}\right],
\label{eq:Pmc}
\end{equation}
with the right-hand side clipped to $[0,1]$, and the prediction of the classical model at temperature $\Theta$ is the Boltzmann average
\begin{equation}
P_{\rm cl}(t;\Theta)=\frac{\int_0^U\!\dd E\; e^{-E/k_B\Theta}\,P_{\rm cl}(t;E)}{\int_0^U\!\dd E\; e^{-E/k_B\Theta}}\,.
\label{eq:Pclth}
\end{equation}

In the semiclassical limit of a deep trap the classical model must become exact, and the correspondence is controlled. The Wigner function of an untruncated thermal state of occupation $\bar n$ is the classical Gaussian at the effective temperature $k_B\Theta_{\rm eff}=(\bar n+\tfrac12)\hbar\omega$, and free flight transports Wigner functions along classical trajectories exactly, the Hamiltonian being quadratic; for the truncated ensembles plotted here the correspondence is semiclassical rather than exact (a six-level renormalized state is no longer Gaussian), with the residual quantum effects residing in the discreteness of the ensemble and in the difference between the bound-state projector and the sharp classical energy shell. Figure~\ref{fig:rr}(b) compares the Boltzmann-weighted quantum signal $\bar P(t)=\sum_{n<N}p_n\,P_{1\mathrm d}(t)$ with Eq.~\eqref{eq:Pclth} at equal temperature: for $k_B\Theta\gtrsim1.5\,\hbar\omega$ the two agree to within $0.015$ at all times, while at $k_B\Theta=0.5\,\hbar\omega$ they separate qualitatively, the classical curve overestimating the survival throughout (maximum deviation $0.08$). The origin of the low-temperature failure is visible in the kernel~\eqref{eq:Pmc}: it is pinned at unity up to the escape time $t_*(E)=(U-E)/\sqrt{UE}$, which diverges as $E\to0$ (a classical particle at rest at the trap center never escapes), so a cold classical ensemble barely decays, whereas the quantum ground state acquires population outside the truncated bound subspace from the outset. Fitting the quantum signals with the classical model quantifies the resulting thermometry bias and confirms the $(\bar n+\tfrac12)$ correspondence: the best classical fit to the $k_B\Theta=0.5\,\hbar\omega$ ensemble returns $k_B\Theta_{\rm fit}\approx0.71\,\hbar\omega$, close to $k_B\Theta_{\rm eff}=0.66\,\hbar\omega$ and a $40\%$ overestimate of the true temperature, and even the pure ground state is well fit by a spurious $k_B\Theta_{\rm fit}\approx0.55\,\hbar\omega$: the classical analysis misreads zero-point motion as thermal energy. A quantum treatment such as Eqs.~\eqref{eq:psin}, \eqref{eq:P1d}, and \eqref{eq:3dfact} is therefore required for atoms prepared near the motional ground state. A quantum description of recapture was previously developed for anti-trapped Rydberg states~\cite{deKeijzer2023}; the present treatment concerns the ground-manifold dynamics relevant to the diagnostic.

Beyond single atoms, the factorized overlaps also determine multi-atom recapture statistics. Consider $N$ noninteracting spin-polarized fermions initially occupying trap eigenstates $\alpha_1,\dots,\alpha_N$ (Pauli exclusion enforces single occupancy). During the dark time each single-particle orbital evolves by Eq.~\eqref{eq:psin}, and the amplitude to recapture the ensemble into the bound Slater determinant $\{\beta_1<\dots<\beta_N\}$ is the determinant of single-particle overlaps $C_{\beta\alpha}(t)=\langle\phi_\beta|\psi_\alpha(t)\rangle$. The probability to recapture all $N$ atoms is
\begin{equation}
P_{N\to N}(t)=\sum_{\beta_1<\dots<\beta_N}^{E_{\bm\beta}<U}
\big|\det\big[C_{\beta_i\alpha_j}(t)\big]\big|^2 ,
\label{eq:PNN}
\end{equation}
which for $N=2$ reads $P_{2\to2}=\sum_{l<m}|C_{l\alpha_1}C_{m\alpha_2}-C_{m\alpha_1}C_{l\alpha_2}|^2$: the exchange term interferes with the direct term. Free propagation preserves parity along each axis [Eq.~\eqref{eq:psin} gives $\psi_n(t,-x)=(-1)^n\psi_n(t,x)$], so $C_{\beta\alpha}$ vanishes unless $\beta$ and $\alpha$ have equal parity along every axis. Two-particle interference therefore requires the two atoms' quantum numbers to match in parity axis by axis, a suppression of order $2^3$ for a generic 3D thermal pair, but only $2$ for an elongated trap in which both atoms occupy the radial ground state. Since the different $C_{\beta\alpha}(t)$ decay on different time scales, the interference is most visible at early times and low temperatures. Finally, the echo protocol of Sec.~\ref{sec:echo} applies verbatim to any atom number: the cycle unitary is $U_{\rm cyc}^{\otimes N}=e^{-i\tau\sum_k H_0^{(k)}/\hbar}$, so an $N$-atom Slater determinant (or any correlated motional state) is returned exactly.

\section{Numerical methods}
\label{app:methods}

Wave-packet dynamics were computed with a Strang split-step Fourier method on grids of $N=2048$--$4096$ points spanning $|x|\le30$--$45\,a_{\rm ho}$, with time step $\dd t=2$--$5\times10^{-4}\,\omega^{-1}$ during trap-on segments; trap-off segments were propagated exactly with the free kinetic propagator. For the Gaussian-trap simulations a complex absorbing potential was applied beyond $0.8$ of the box edge, and initial ground states were obtained by imaginary-time propagation; energies were evaluated spectrally. Atom loss was quantified by projecting the final wave packet onto the negative-energy eigenstates of the one-dimensional Gaussian trap, obtained by exact diagonalization (Fourier-grid and finite-difference Hamiltonians); the bound-state count converges to 378 under joint refinement of the box and grid, and the finite-box projector used for the quoted numbers contains 376 states. The result is a conservative, converged numerical estimate for the one-dimensional radial model (not a three-dimensional tweezer-loss calculation), insensitive to box, grid, and time step; the absorbing potential remains inactive at the reported operating points. The reported residuals are $\Delta n=\langle H\rangle/\hbar\omega-\langle H\rangle_0/\hbar\omega$ with $H$ the final (nominal-depth) trap Hamiltonian. Ermakov trajectories [Eq.~\eqref{eq:ermb}] were integrated with an eighth-order Runge--Kutta method at relative tolerance $10^{-12}$. Multimode echoes were evaluated by exact $2\times2$ symplectic products; roots and continuation branches were polished by bounded least squares, and local regularity was evaluated from analytic endpoint Jacobians. The shortest-root statement within the two-pulse palindrome was certified by the exact rational reduction of Eqs.~\eqref{eq:mm_tangent_recurrence_app}--\eqref{eq:mm_rational_closure_app} and by a Sturm-sequence root count performed with exact integer-polynomial arithmetic. Longer alternating words were tested by constrained multistart minimization. The time-optimality search of Sec.~\ref{sec:optimal} used sequential quadratic programming over the segment durations of $\mathrm{on}\,\mathrm{off}\,\mathrm{on}\,\mathrm{off}$ sequences with the endpoint $(b,\dot b)=(1,0)$ imposed as an equality constraint. Three hundred random initial duration vectors were used at each operating point, more than $88\%$ of which converged, and none produced a recovery shorter than the analytic two-switch solution.

\bibliography{refs}

\end{document}